\documentclass[journal]{IEEEtran}

\usepackage[cmex10]{amsmath}

\usepackage[square,sort,comma,numbers]{natbib}
\usepackage[]{graphicx,color}
\usepackage[]{subfigure}

\begin{document}
%
\title{A new Level-set based Protocol for Accurate Bone Segmentation from CT Imaging}

\author{Manuel~Pinheiro~and~J.~L.~Alves
\thanks{The authors are with the Department of Mechanical Engineering, University of Minho, 4800-058 Guimar\~{a}es, Portugal (e-mail: manuelspinheiro@gmail.com; jlalves@dem.uminho.pt)}
\thanks{}}

\maketitle

\begin{abstract}
In this work it is proposed a medical image segmentation pipeline for accurate bone segmentation from CT imaging. It is a two-step methodology, with a pre-segmentation step and a segmentation refinement step. First, the user performs a rough segmenting of the desired region of interest. Next, a fully automatic refinement step is applied to the pre-segmented data. The automatic segmentation refinement is composed by several sub-stpng, namely image deconvolution, image cropping and interpolation. The user-defined pre-segmentation is then refined over the deconvolved, cropped, and up-sampled version of the image. The algorithm is applied in the segmentation of CT images of a composite femur bone, reconstructed with different reconstruction protocols. Segmentation outcomes are validated against a gold standard model obtained with coordinate measuring machine Nikon Metris LK V20 with a digital line scanner LC60-D that guarantees an accuracy of 28 $\mu m$. High sub-pixel accuracy models were obtained for all tested Datasets. The algorithm is able to produce high quality segmentation of the composite femur regardless of the surface meshing strategy used.
\end{abstract}


\IEEEpeerreviewmaketitle

\section{Introduction}

\IEEEPARstart{T}{he} first milestone towards custom implant development is the accurate extraction of the target anatomical structure from medical image data. The tremendous evolution of CT imaging led to the widespread of this technique to all medical fields. From the engineering standpoint, CT imaging can be used for the development of patient-specific biomechanical and finite element models, as well as in the development of custom implants \citep{bargar1989shape}\citep{garg1985design}\citep{stulberg1989rationale}.

Currently, CT imaging is the modality of choice for imaging the Human skeletal system. The ability to enhance the radiological contrast between hard and soft-tissue facilitates image segmentation, and the production of accurate representations of bone. The accurate segmentation of bone is the only way to guarantee the overall fit to the patient's anatomy, which may be paramount for success of the implant \citep{engh1985biological}\citep{o1987bone}. Nevertheless, the degree of patient fit necessary to minimize the biological impact of the implant is still unknown. On the one hand, too much implant fit to the target anatomy may preclude implant insertion and may cause severe damages to the host bone \citep{walker1988design}. On the other hand, the absence of implant fit may cause interfacial micromotions that prevent bone ingrowth and implant osseointegration \citep{walker1987strains}\citep{mandell2004conical}.

During image acquisition the CT scanner acts as a low-pass filter, and there is a degradation in image quality due to the limited frequency response of the imaging system. The sharp transitions between adjacent structures found in reality become diffuse in the final image. In \citep{hangartner1996evaluation} it was concluded that CT imaging produced large domain overestimations for structures with a cortical thickness below 2.0 mm. The amount of spatial image blur is often modelled in the image space by the system's Point Spread Function (PSF). In addition, the inner and the outer cortical surfaces of bone could only be accurately determined for thickness greater than the Full Width at Half Maximum (FWHM) of the PSF \citep{prevrhal1999accuracy}. For objects with a thickness smaller than the FWHM of the PSF, domain overestimations up to 40\% of the original size were obtained in \citep{kang2003new}. Similarly, in \citep{ohkubo2008imaging} it was observed that for small diameter spheres imaged with CT the apparent diameter was in fact the FHWM of the PSF. Therefore, the FWHM seems to provide a measure of the maximum spatial frequency that can be accurately encoded by a given CT machine, rather than other acquisition settings. 

Another limiting factor to image accuracy may be the reconstruction (Field of View) FOV. In high-resolution reconstructions the PSF if often the limiting factor, however for larger FOV if the pixel size is larger than the FWHM, the PSF is spread to occupy a single pixel in the reconstructed image \citep{dougherty2009digital}. Image reconstruction with large voxel sizes has a higher impact on bone thickness overestimation than smaller sizes, and that large voxel sizes were also detrimental for the representation of thinner and highly curved structures \citep{maloul2011impact}.

Although relevant, the limitation of the image acquisition process may be minimal considering other error sources. In fact, the image reconstruction in CT imaging is known to have a very high accuracy, and to be almost free of geometrical magnification \citep{hildebolt1990validation}. Image segmentation is often affected by high inter and intra-expert variability, and the image processing and segmentation chain may contribute up to 70\% of the average error found in the final reconstruction \citep{wang2009precision}.

In the literature a plethora of studies have addressed bone segmentation, however few have evaluated the accuracy of the segmentation outcome. In an early study \citep{rothuizen1987accuracy} analysed the CT attenuation profile normal to the bone's surface, and concluded that a single threshold was insufficient to accurately define the femur's cortical shell. The relative thresholds of 45\% and 50\% of the maximum HU profile value were proposed to segment properly the diaphyseal and metaphyseal regions of the femur, respectively. More recently, \citep{kang2003new}, \citep{aamodt1999determination} and \citep{rathnayaka2011effects} reported the accurate segmentation of cortical bone from CT imaging with single thresholding, adaptive thresholding, and multiple thresholding. Exploiting the concept of relative thresholding, in \citep{treece2010high} a Levenberg-Marquardt segmentation algorithm to quantify femoral cortical thickness with sub-millimetre accuracy was proposed. An ideal high-resolution attenuation model was fitted to the HU attenuation profile normal to the bone's surface and then thresholded with the 50\% relative threshold. A similar approach was used in \citep{pakdel2012generalized} to segment bones from the craniofacial skeleton. The production of sub-pixel accuracy estimates of both inner and outer surfaces of the bone were reported.

In practice, the application of the aforementioned techniques may possess some limitations. Single thresholding are very sensitive to image inhomogenities, noise, and threshold selection. With adaptive thresholding there is no guarantee that a closed contour will be obtained \citep{kang2003new}, and may provide incorrect estimates of the bone surface especially in regions with thin cortical shells \citep{prevrhal1999accuracy}. Relatively to model fitting techniques, the application of adaptive/relative thresholds to each boundary pixel is extremely time consuming and unpractical. In addition, the normal direction is highly affected by the discrete nature of the image \citep{yao2005estimation}. The surface of the bone has to be sampled and each surface point needs to be processed independently, which produces highly irregular contours. It also produces unreliable estimates of the bone surface if the observed attenuation profile deviates from the ideal attenuation profile, which is common near the articulating surfaces of bones.

In this work a different approach to accurate bone segmentation is proposed. The proposed protocol allows the segmentation of bony structures with sub-pixel accuracy, and intrinsically guarantees the smoothness of the extracted contours. The proposed segmentation protocol is validated through the comparison between the segmentation outcome and a geometrically well-defined gold standard. For validation purposes a synthetic bone was used as the gold standard. The remainder of this paper is organized as follows: in section \ref{secII} the segmentation protocol, the definition of the segmentation gold standard, and the means for quantifying the accuracy of the CT machine are described; in section \ref{secIII} one presents a methodology to estimate the PSF of the CT machine; in section \ref{secIV} the impact of the domain discretization (voxel vs. average error) in the model accuracy is shortly analysed; section \ref{secV} and section \ref{secVI} refer to the description and discussion of the results obtained with the proposed segmentation pipeline; and ultimately in section \ref{secVII} one presents the conclusions of the present work.

\section{Materials and Methods \label{secII}}

\begin{figure}[htb]
	\centering
	\subfigure[]{
	      \label{fig:ch03fig01_1}
	      \includegraphics[width = 88.0 mm]{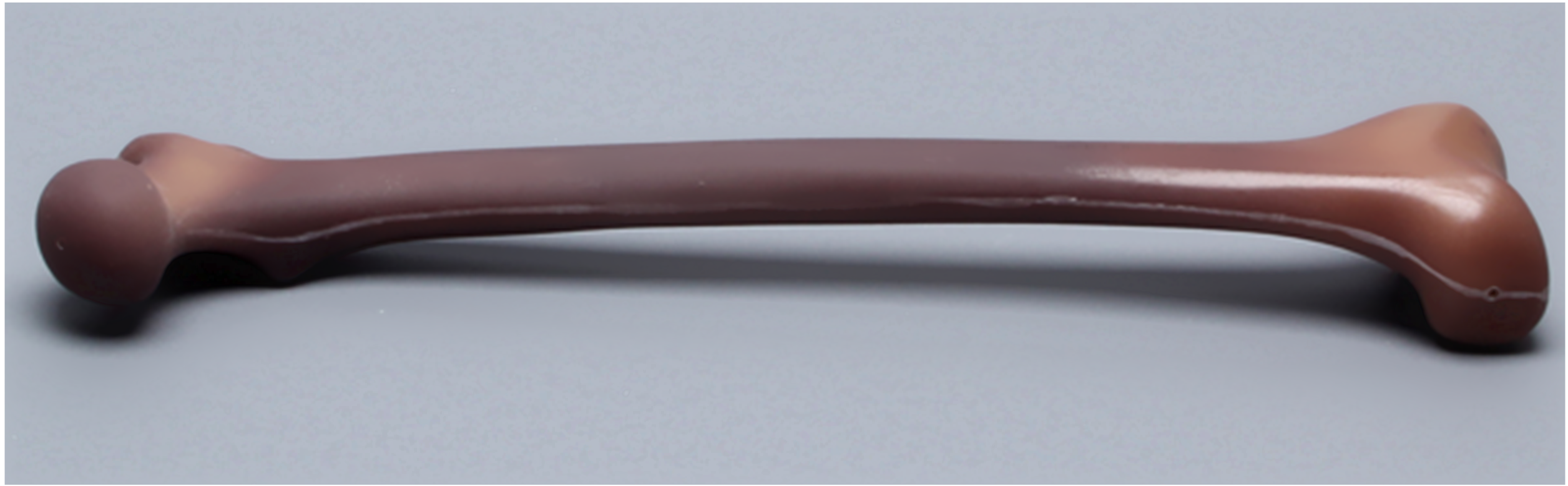}}
	\subfigure[]{
	      \label{fig:ch03fig01_2}
	      \includegraphics[width = 27.0 mm]{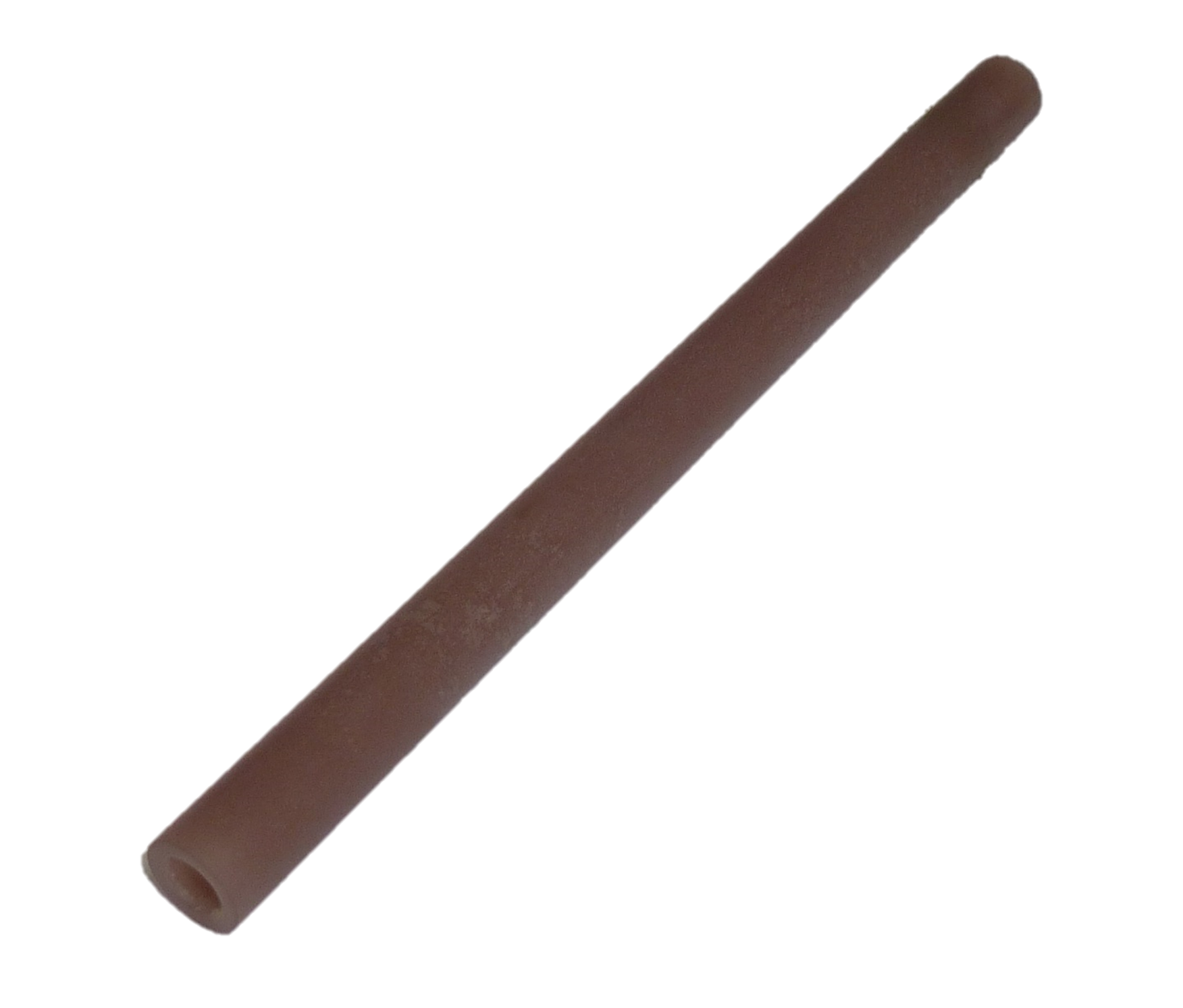}}
	\subfigure[]{
	      \label{fig:ch03fig01_3}
	      \includegraphics[width = 27.0 mm]{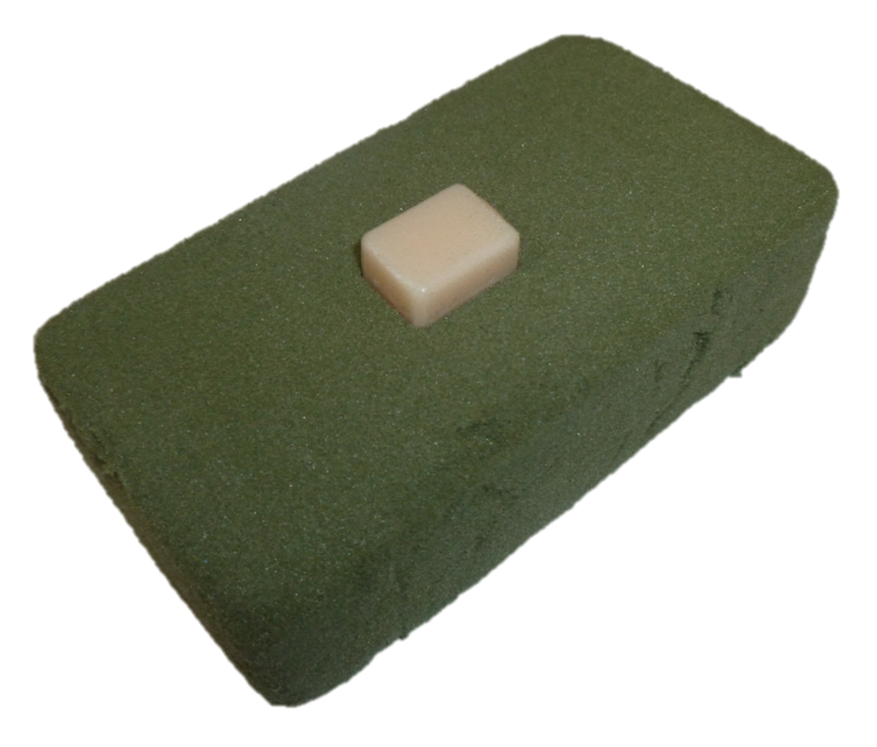}}
	\subfigure[]{
	      \label{fig:ch03fig01_4}
	      \includegraphics[width = 27.0 mm]{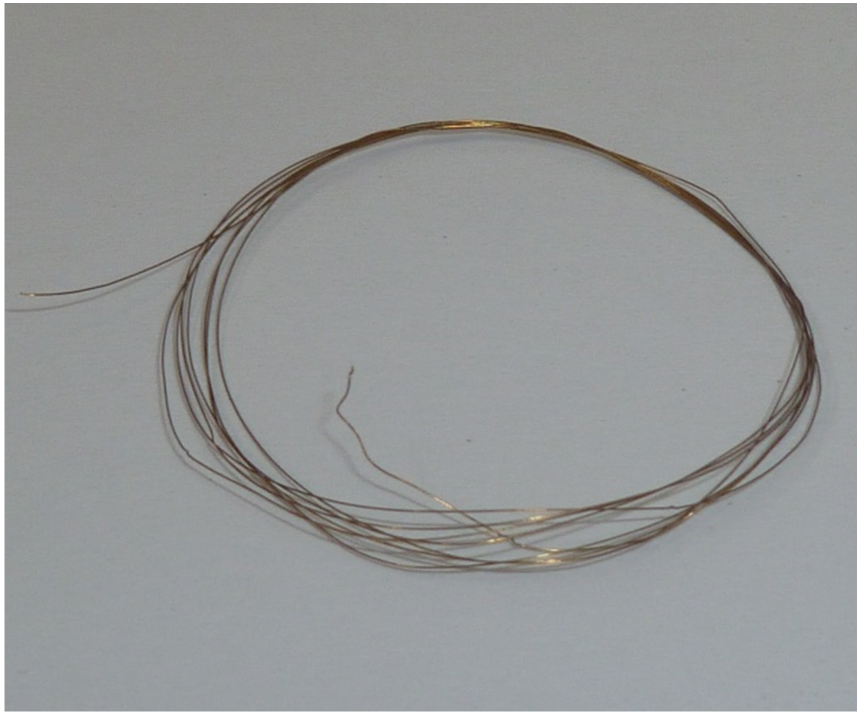}}					
	\caption{In (a) the gold standard composite femur and in (b), (c) and (d) the reference calibrated hollow cylinder with $9.81$ mm of diameter, the ceramic box and the brass alloy wire with $0.10$ mm of diameter used to estimate the system's Point Spread Function}
	\label{fig:ch03fig01}
\end{figure}

Due to the degree of variability in the Human anatomy and image artifacts, segmentation methods designed specifically to each application often produce better results than general purpose algorithms. Nevertheless, some degree of standardization is desirable, particularly when the segmentation is part of product development pipeline. Therefore, in this work one proposes a two-step segmentation protocol for reliably and accurately extract hard tissue structures from image data. To evaluate the accuracy of the newly proposed segmentation protocol a phantom study was carried out. One composite replica of the Human femur (Fig. \ref{fig:ch03fig01}) commercially available at the Sawbones\footnote{http://www.sawbones.com/} was imaged with a CT machine. The image acquisition process was carried with a fourth-generation CT scanner Toshiba $Aquilion^{TM}$ $64$ at the CUF Hospital, Porto (Portugal).

During image acquisition and reconstruction two types of scan settings were used. Three geometrically simpler phantom objects were added to the acquisition process. In practice, these geometrically simpler phantoms may be important to characterize the image acquisition process, especially the limiting resolution of the CT scanner. The tested objects were a calibrated hollow cylinder with an outer diameter of $ 9.81 \pm 0.02 $ mm, a brass alloy wire with 0.10 mm of diameter, and a ceramic box with dimensions $12.51 \times 13.81 \times 18.01 \pm 0.01$ mm (Fig. \ref{fig:ch03fig01} (b), (c) and (d)). The compact nature of these phantom objects allow them to be imaged simultaneously with the patient, and may avoiding the need to have a dedicated phantom to assess the spatial resolution of the CT scanner. The ability to quantify the system's limiting resolution was evaluate against the CATPHAN 528, which is a commercially available phantom often used for quality control. The CATPHAN 528 was imaged to quantify the true in-plane resolution of the CT machine, against which the performance of each phantom could be compared. The summary of image acquisition protocols and image reconstruction resolutions for the phantom femur, the CATPHAN 528, and the three other phantoms is presented in Table \ref{tab:ch03Tab01}. 

\begin{table*}[t]
\linespread{1.0}\selectfont
	\centering
	\caption{CT image acquisition protocol summary and target reconstruction resolution for each Dataset (DS): Datasets \#1, \#2 and \#3 were obtained from the raw data, whereas in Datasets \#4 and \#5 the in-plane resolution was downscaled to $1/2$ and $1/4$ of the original (reconstructed) resolution}
	\resizebox{181.0 mm}{!}{
	\begin{tabular}{lccc|cccc|cc} \hline \hline
				& \multicolumn{9}{c}{Data Acquisition} \\		\hline 
		\multicolumn{1}{l}{}  & \multicolumn{1}{c}{DS\#1} & \multicolumn{1}{c}{DS\#2} & \multicolumn{1}{c|}{DS\#3} & \multicolumn{1}{c}{Cylinder} & Box
		& \multicolumn{1}{c}{CATPHAN 528} & Wire  & DS\#4 & DS\#5 \\ \hline
		Tube Voltage, $kV$		& 120 		& 120			& 120			& 120		& 120		& 120			& 120		& 120			& 120     \\
		Tube Current, $mA$		& 200			& 200			& 200			& 200		& 200		& 200			& 200		& 200	 		& 200     \\
		Scan FOV, $mm$	    	& 240			& 400			& 400			& 240		& 400		& 400			& 400		& 240			& 240     \\
		Slice Thickness, $mm$ & 0.3			& 0.3			& 3.0			& 0.3		& 0.3		& 0.3			& 0.3		& 0.5			& 1.0	    \\
		Pixel Spacing, $mm$   & 0.243		& 0.525		& 0.525		& 0.243	& 0.525 & 0.460		& 0.460 & 0.486		& 0.972	  \\
		Reconstruction Kernel & FC 30		& FC 84		& FC 84		& FC 30	& FC 84 & FC 84		& FC 84 & FC 30		& FC 30	  \\
		Image Matrix 					& 512x512	& 512x512	& 512x512	& 			&       & 512x512	&       & 256x256	& 128x128 \\ \hline \hline
	\end{tabular}}
	\label{tab:ch03Tab01}
\end{table*}

Regarding the segmentation protocol, a pre-segmentation of the image $I(x,y)$ must be provided. In this step the technique(s) more suitable to provide an initial segmentation of the bone may be applied. The output of image pre-segmentation should contain a set of Regions of Interest (ROIs) (for instance, the segmentation of the composite femur and the phantom object(s)), and may also have some additional spatial constraints, such as the CT table. The pre-segmentation aims to provide some high-level information about the desired domain, as well as some spatial relationships between any existing adjacent structures. This allows us to handle segmentation variability prior to segmentation refinement. In the second step (which will be referred as refinement step), a fully automatic segmentation refinement composed by several sub-stpng is performed in order to optimize the initial partition. Fig. \ref{fig:ch03fig02} schematically depicts the proposed segmentation protocol. The refinement protocol comprises image deconvolution, which is applied in order to minimize the partial volume effect caused by the PSF during image acquisition. Next, the image is cropped around the ROI according to the pre-segmented data, and up-sampled with cubic spline interpolation (\ref{e:ch03Eq01}):

\begin{figure}[htb]
	\begin{center}
		\includegraphics[width=88.0 mm]{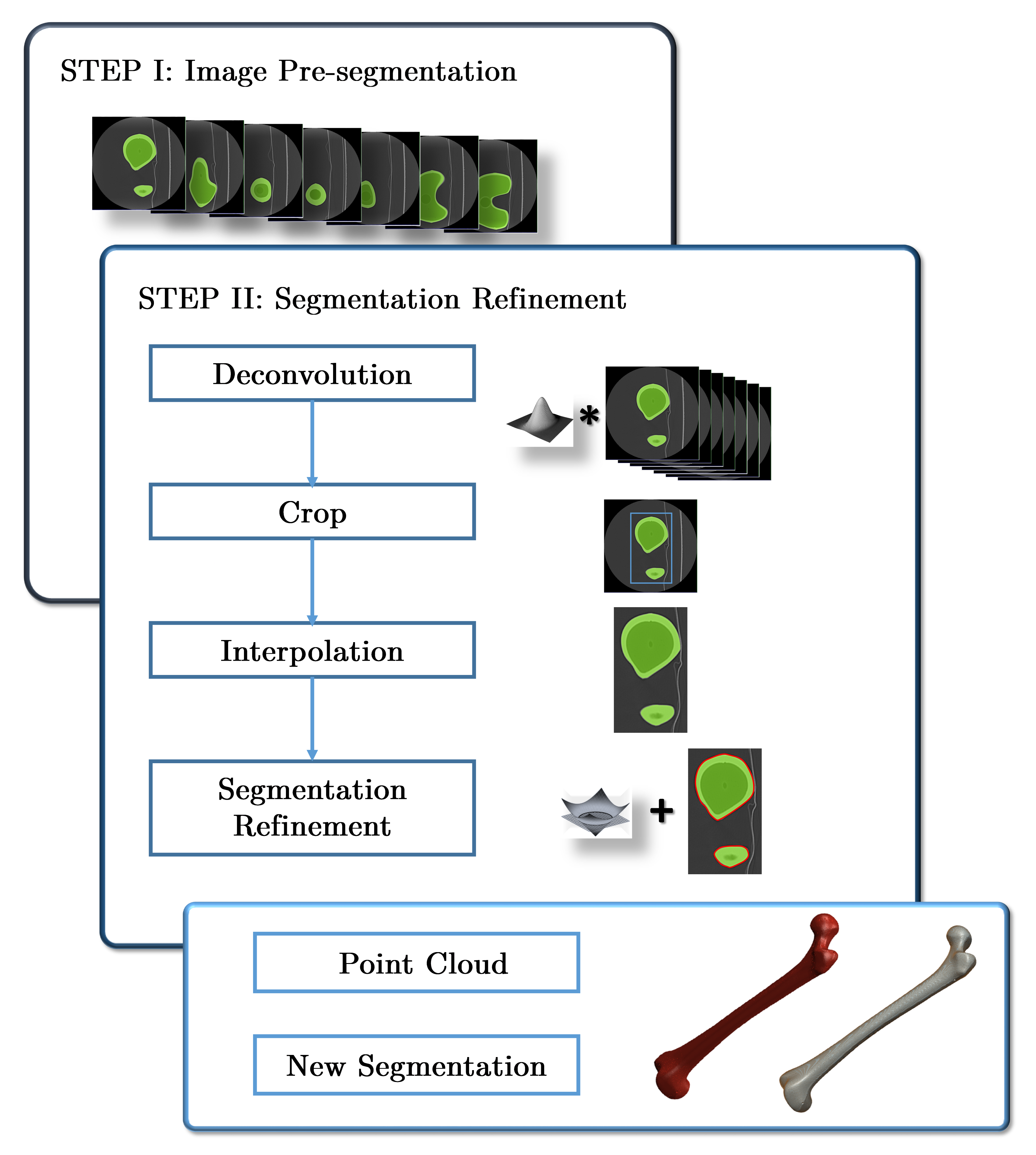}				
		\caption{Schematic description of the image segmentation protocol proposed: in a first step the user performs a pre-segmentation of the domain that aims to provide some high level information of the desired ROI and adjacent structures; in a second step the pre-segmentation provides the starting point for a fully automatic segmentation refinement which encompasses image deconvolution, image cropping, interpolation and segmentation by the level set method}
		\label{fig:ch03fig02}
	\end{center}
\end{figure}

\begin{equation} \label{e:ch03Eq01}
	\beta ^3  = \left\{ \,
		\begin{IEEEeqnarraybox}[][c]{l?s}
		\IEEEstrut
		\frac{2}{3} - \frac{1}{2}\left| x \right|^2 \left( {2 - \left| x \right|} \right),\ \ 0 \le \left| x \right| < 1 \\
		\frac{1}{6}\left( {2 - \left| x \right|} \right)^3 ,\                               \ 1 \le \left| x \right| < 2 \\
    0,\                                                                                 \ 2 \le \left| x \right|
		\IEEEstrut
	\end{IEEEeqnarraybox} \right.
\end{equation}

where $x$ defines the finite support of the basis function. In \citep{thevenaz2000interpolation} concluded that cubic spline interpolation provides the best interpolation strategy for image processing applications, both in terms of computation effort and image induced artifacts. Segmentation refinement is performed over the interpolated ROI. The final segmentation obtained using the Chan-Vese level-set method, with the additional spatial constraints imposed in the pre-segmentation step (\ref{e:ch03Eq02}):

\begin{IEEEeqnarray}{rCl} \label{e:ch03Eq02}
\frac{{\partial \phi _i }}{{\partial t}} & = & \tau div\left( {\frac{{\nabla \phi _i }}{{\left| {\nabla \phi _i } \right|}}} \right) - \lambda _1 \left( {I - \mu _1 } \right)^2 + \lambda _2 \left( {I - \mu _2 } \right)^2 \nonumber \\
&& + \eta \mu _2 \sum\nolimits_{k = 1;k \ne i}^n {S{}_k}
\end{IEEEeqnarray}

where $\phi_i$ is the level-set corresponding to the pre-segmented region $S_i$, $\lambda _1 = \lambda _2 = 1$, $\mu _1$ and $\mu _2$ are the average intensity inside and outside the curve, and the final summation assigns the average intensity of the background to all the remaining pre-segmented sub-regions $S_k$ multiplied by a user defined cost $\eta$.

The Chan-Vese ACWE level-set method is less sensitive to curve initialization and noise than other level-set methods, which allows a less restrictive pre-segmentation and a lesser dependency between the pre-segmentation and the final segmentation. Segmentation refinement was performed independently for each region, instead of applying a multi-phase level-set method (see for instance \citep{vese2002multiphase}). This reduces the computational complexity and memory requirements of the segmentation refinement, by applying image interpolation in smaller region at each iteration. Two refinement outputs were considered (Fig. \ref{fig:ch03fig02}). The new segmentation can be directly exported to the CAD modelling software via point cloud or downscaled to the original image resolution for further surface or Finite Element mesh generation. To test the robustness against image noise, the segmentation refinement was tested over Datasets \#1 to \#5 corrupted with Additive White Gaussian Noise (AWGN).  

\section{Image Deconvolution \label{secIII}}

It is known from image acquisition theory that any image obtained from an imaging system is not perfect and is only an approximation to the real (ideal) image. The real image is never available due to the intrinsic nature of the acquisition process. However, an estimate of its true distribution may be obtained considering the output image and some prior knowledge about the system's behaviour. In 2D image acquisition theory, the imaging system is commonly considered as being linear and spatially invariant, and the output image $G(x,y)$ may be correlated with the input image $I(x,y)$ according to (\ref{e:ch03Eq03}):

\begin{IEEEeqnarray}{c} \label{e:ch03Eq03}
	G\left( {x,y} \right) = I\left( {x,y} \right) \otimes h\left( {x,y} \right) + n\left( {x,y} \right) 
\end{IEEEeqnarray}

where $\otimes$ denotes the 2D convolution, $h(x,y)$ denotes the system blurring effect of the system's PSF, and $n(x,y)$ is an addictive noise term \citep{campisi2007blind}. The deconvolution problem is intrinsically limited by the knowledge about the PSF and the noisy processes related with the acquisition itself. Image restoration in this work is accomplished by applying a standard iterative blind deconvolution algorithm. The MATLAB function \emph{deconvblind} is used for the purpose of image restoration. The \emph{deconvblind} uses the Lucy-Richardson algorithm to obtain the new estimates of the original scene $\hat I_{k + 1} \left( {x,y} \right)$ and the PSF $\hat h_{k + 1} \left( {x,y} \right)$ and can be defined as (\ref{e:ch03Eq04}):

\begin{IEEEeqnarray}{llrllr} \label{e:ch03Eq04}
	\IEEEyesnumber \IEEEyessubnumber*
	\hat I_{k + 1} \left( {x,y} \right) & = & \\ \nonumber
& \hat I_k \left( {x,y} \right) \left[ {\frac{{G\left( {x,y} \right)}}{{\hat h_k \left( {x,y} \right) \otimes \hat I_{k + 1} \left( {x,y} \right)}} \otimes \hat h_k^* \left( {x,y} \right)} \right]\\				
		\hat h_{k + 1} \left( {x,y} \right) & = &\\ \nonumber
		& \hat h_k \left( {x,y} \right) \left[ {\frac{{G\left( {x,y} \right)}}{{\hat h_k \left( {x,y} \right) \otimes \hat I_{k + 1} \left( {x,y} \right)}} \otimes \hat I_k^* \left( {x,y} \right)} \right]	
\end{IEEEeqnarray}

where $\hat h_k^* \left( {x,y} \right)$ and $\hat I_k^* \left( {x,y} \right)$ are the complex conjugates of $\hat h_{k} \left( {x,y} \right)$ and $\hat I_{k} \left( {x,y} \right)$, respectively, and where $k \in N_0$, and $\hat I_{0} \left({x,y} \right) = G(x,y)$ is the acquired image and the image noise n(x,y) is neglected. The blind deconvolution is not dependent on the knowledge of the system's spatial blurring \citep{richardson1972bayesian}\citep{ayers1988iterative}\citep{solomon2011fundamentals}\citep{pantin2007deconvolution}. However, we found that robust results may be obtained if an accurate initial guess of the system's PSF is provided. The PSF is determined by the overall behaviour of the image acquisition system. The noise term $n(x,y)$ is typically a stochastic process that may be originated by a multitude of processes \citep{solomon2011fundamentals}. Noise can be efficiently suppressed by non-linear filtering, for instance applying anisotropic diffusion proposed in \citep{perona1990scale}. It also avoids edge bias, which is very common with linear filtering \citep{verbeek1994location}\citep{mendoncca2004bias}\citep{bouma2005correction}.

In practice, the PSF of a given imaging system is frequently approximated by a normalized Gaussian function as stated by the central slice theorem, such that (\ref{e:ch03Eq05}):

\begin{IEEEeqnarray}{c} \label{e:ch03Eq05}
	h\left( {x,y,z} \right) = \frac{1}{{\left( {2\pi } \right)^{\frac{3}{2}} \sigma _x \sigma _y \sigma _z }}e^{\left( {\frac{{x^2 }}{{2\sigma _x }} + \frac{{y^2 }}{{2\sigma _y }} + \frac{{z^2 }}{{2\sigma _z }}} \right)}
\end{IEEEeqnarray}

where $\sigma _x$, $\sigma _y$, and $\sigma _z$ denote the standard deviation in each orthogonal direction. Two assumptions are commonly found in literature for the PSF namely: (i) the PSF is assumed to be uniformly invariant in the slice plane; and (ii) the cross-plane PSF is generally also assumed to be invariant in the axial direction \citep{prevrhal1999accuracy}\citep{treece2010high}\citep{easton2010fourier}\citep{pakdel2012generalized}. In practice, the PSF is not completely isotropic and shift invariant, however this approximation can be safely made for most CT scanners, as well as for several other medical acquisition modalities \citep{nickoloff1985simplified}\citep{dore1997experimental}. For simplicity one will only consider the estimation of the in-plane blur, and hence (\ref{e:ch03Eq05}) can be reduced to its 2D counterpart.

\begin{figure}[htb]
	\begin{center}
		\includegraphics[width=88.0 mm]{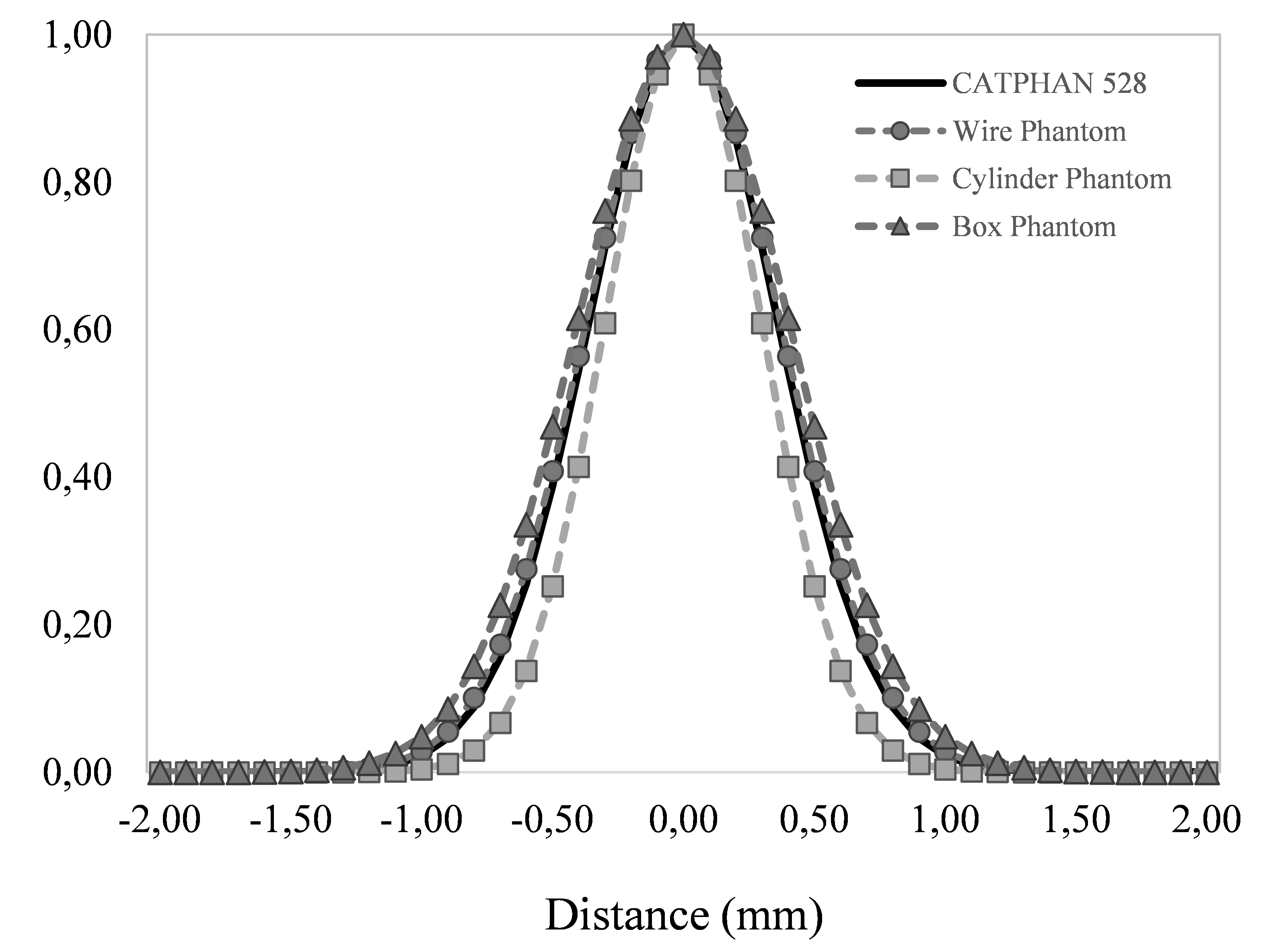}				
		\caption{The PSF of the fourth-generation Toshiba Aquilion$^{TM}$ 64 CT scanner with the CATPHAN 528, the $0.10$ mm wire phantom, and the calibrated hollow cylinder and ceramic box phantoms}
		\label{fig:ch03fig04}
	\end{center}
\end{figure}

There are several approaches to determine the PSF of an imaging system proposed in the literature. The phantom objects investigated in this work allow the measurement of PSF directly by simulating an impulse signal, or by computing the Edge Spread Function (ESF) along the cylinder and the ceramic box. The ESF can be computed simply by differentiating the edge response to radiopaque objects \citep{cunningham1987method}\citep{cunningham1992signal}\citep{samei1998method}\citep{mori2009deriving}. In the ESF a Gaussian function was fitted to the gradient of the attenuation profile normal to each object' surface. For profile de-noising, Wavelet thresholding was applied to the 1D signals as proposed in \citep{donoho1995adapting}\citep{donoho1995noising}. Curve fitting was performed over a narrow band around the gradient maximum, in order to remove the influence of the adjacent structures \citep{smith1997scientist}. A similar approach was applied in \citep{joshi2008psf} to completely characterize the PSF using a single image. The final estimate of the PSF was defined as the average of all PSF estimates along the sampled phantom edges. Fig. \ref{fig:ch03fig04} depicts the PSF of the Toshiba $Aquilion^{TM}$ $64$ CT scanner obtained directly from the CATPHAN 528 and wire phantom, as well as the estimates obtained through the ESF.

\section{Discretization in Domain Accuracy \label{secIV}}

The system's PSF may not be the only factor affecting the resolution of the CT scan. Image reconstruction with large slice thicknesses is commonly associated with larger model inaccuracies \citep{prevrhal1999accuracy}\citep{sato2003limits}\citep{goto2007accuracy}. Nevertheless, in a recent study \citep{meurer2013influence} it was demonstrated that alongside with the slice thickness, the FOV also influences the spatial resolution of the scan and the amount of partial volume averaging. The size of the voxel may be more critical than other potential error sources, such as the surface meshing parameters. In \citep{cortez20133d} the impact of the voxel size and meshing parameters in the representation of a Human lumbar motion segment was analysed. It was concluded that the resolution of the CT scan (voxel size and slice thickness) was the major source of geometrical inaccuracies of the reconstructed model. Therefore, the effect of domain discretization (slice thickness and voxel size) cannot be neglected when assessing the achievable model accuracy from a given set of CT images. To understand the effect of pixel size in domain accuracy, the gold-standard model for composite femur was discretized with different isotropic voxels sizes, namely $\left\{5.0, 3.0, 2.0, 1.5, 1.0, 0.5, 0.2 \right\}$ (mm).

\begin{figure}[htb]
	\centering
	\subfigure[]{
	      \label{fig:figure4_6_1}
	      \includegraphics[width=88.0 mm]{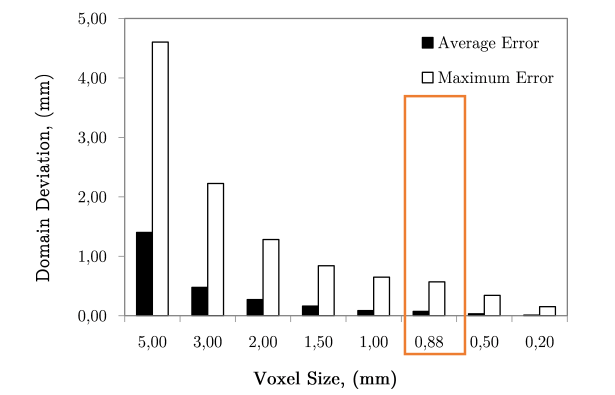}}
	\subfigure[]{
	      \label{fig:figure4_6_2}
	      \includegraphics[width=88.0 mm]{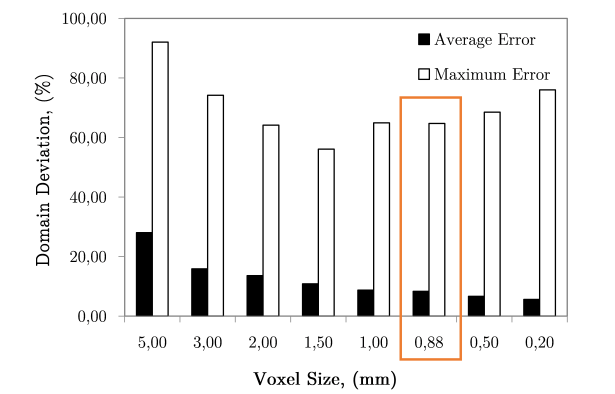}}					
	\caption{In (a) the average and maximum surface errors between the Nikon Metris $LK$ $V20$ gold standard and the surface mesh obtained after domain discretization for all voxel sizes, and in (b) average and maximum deviation normalized by the voxel size for all voxel sizes}
	\label{fig:ch03fig06}
\end{figure}

Fig. \ref{fig:ch03fig06} shows the average and maximum domain error as a function of the voxel size. The average error due to the domain discretization varied between $1.401 \pm 0.836$ mm and $0.011 \pm 0.012$ mm, whereas the maximum deviation was $4.603$ mm and $0.152$ mm for a voxel size of $5.0$ mm and $0.2$ mm, respectively. As expected, the results show that as the domain sampling increases both the absolute average and the maximum error decrease (Fig. \ref{fig:ch03fig06} (a)). The largest domain deviations are found in small localized sharp features or surface irregularities along the reference model, which are lost or smoothed due to the domain sampling. For a voxel size consistent with the size of the FWHM of the PSF ($0.88$ mm), theoretically an average error of $0.073 \pm 0.066$ mm and a maximum deviation of $0.569$ mm are expected to occur due to the domain discretization.

In Fig. \ref{fig:ch03fig06} (b) the average and maximum errors are normalized against the sampling size. The normalized average error decreases from 28\% to 5.6\% of the sampling size, while the maximum error decreases from 92.1\% to 56.1\% for domain samplings between 5.0 mm and 1.5 mm. However, for edge lengths equal and below 1.0 mm the maximum error shows a growth tendency, reaching 76.0\% of the domain sampling for 0.2 mm. The normalized average error decreases and becomes almost constant for smaller sampling sizes, whereas the maximum normalized error increases initially for larger sampling sizes and increases as the sampling size becomes smaller. These provide an evidence that domain deviations may exist regardless of the sampling size used, and that the average deviations converge to a fixed percentage of the sampling size.

\section{Image Segmentation and Refinement \label{secV}}

\begin{figure*}[t]
	\centering
	\subfigure[]{
	      \label{fig:figure4_15_1}
	      \includegraphics[width=44.15 mm]{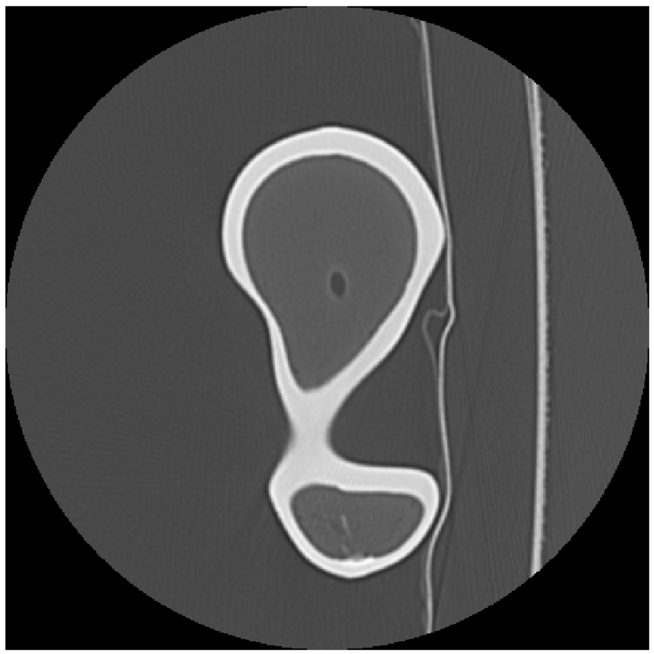}}
	\subfigure[]{
	      \label{fig:figure4_15_2}
	      \includegraphics[width=44.15 mm]{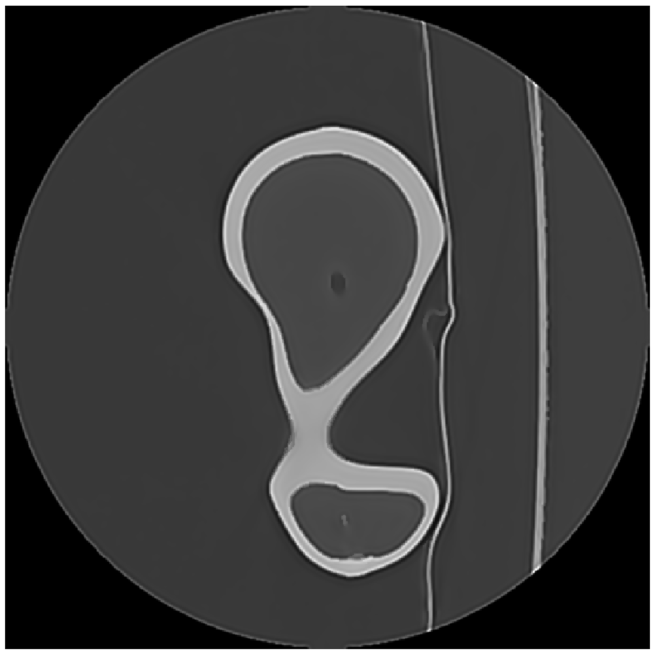}}
	\subfigure[]{
	      \label{fig:figure4_15_3}
	      \includegraphics[width=23.45 mm]{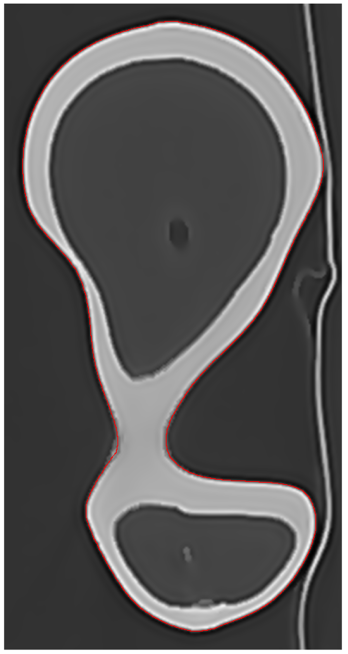}}
	\subfigure[]{
	      \label{fig:figure4_15_4}
	      \includegraphics[width=44.05 mm]{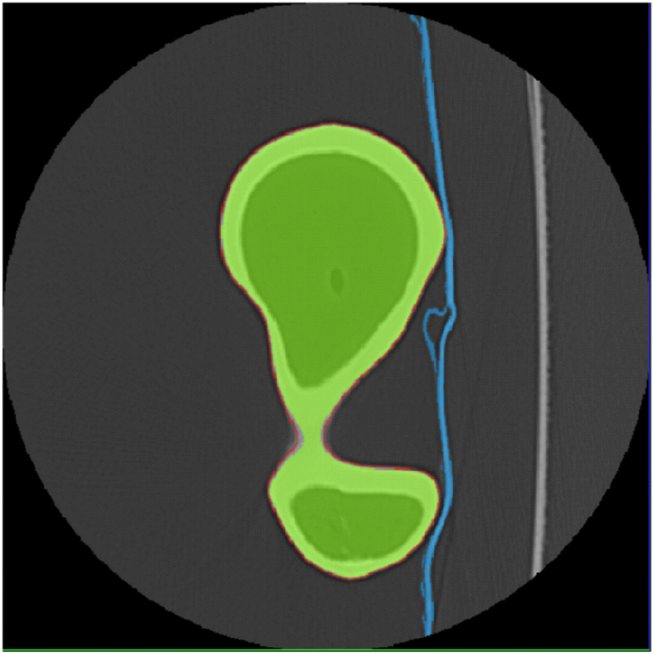}}	
	\subfigure[]{
	      \label{fig:figure4_15_5}
	      \includegraphics[width=44.15 mm]{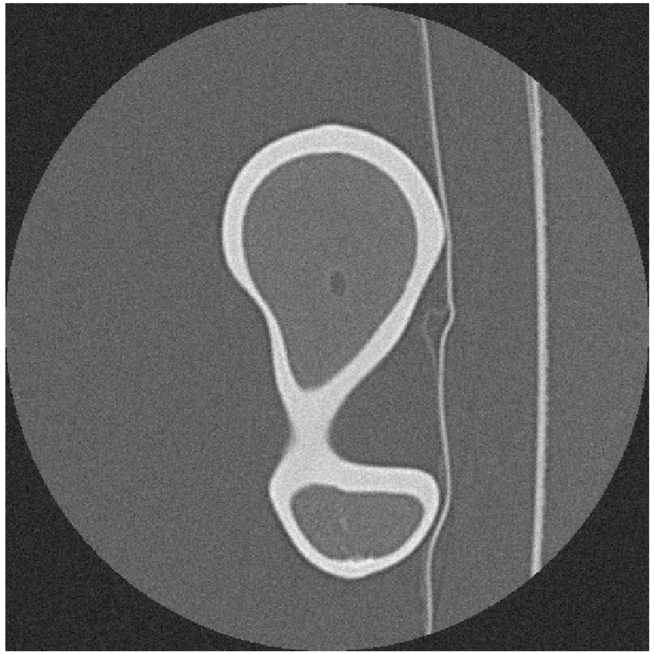}}
	\subfigure[]{
	      \label{fig:figure4_15_6}
	      \includegraphics[width=44.15 mm]{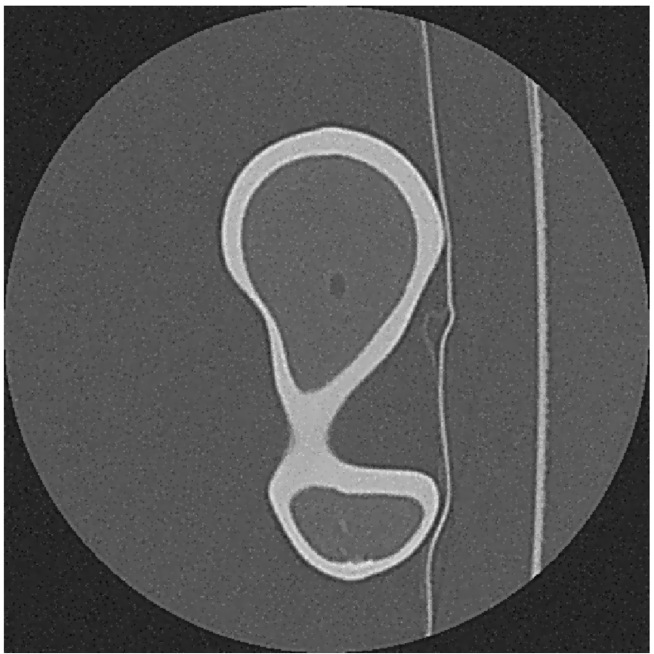}}														
	\subfigure[]{
	      \label{fig:figure4_15_7}
	      \includegraphics[width=23.45 mm]{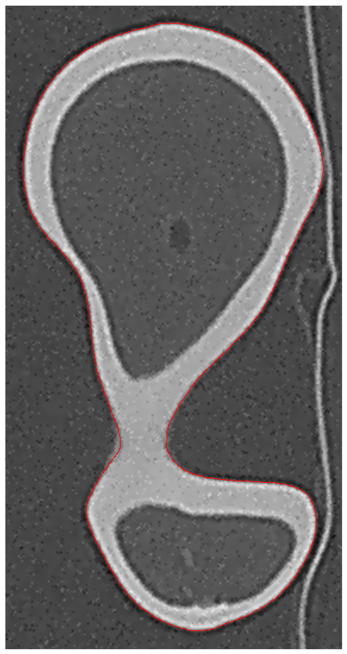}}
	\subfigure[]{
	      \label{fig:figure4_15_8}
	      \includegraphics[width=44.05 mm]{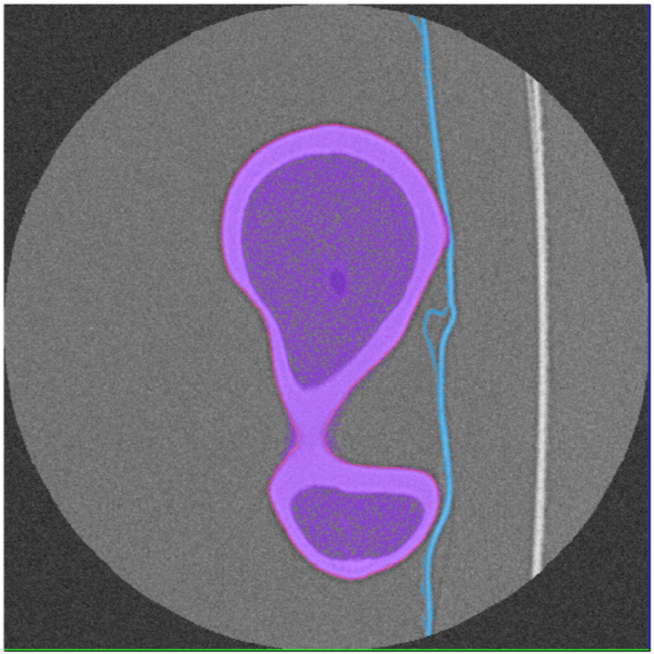}}
	\caption{Segmentation refinement pipeline applied to the Dataset \#1: in (a) the original image data without noise; in (b) the image data after de-noising with anisotropic diffusion and image deconvolution; in (c) image cropping and interpolation around the pre-segmented region; and in panel (d) the final segmentation contour superimposed over the pre-segmentation mask; in (e), (f), (g) and (h) image refinement is applied to the same image corrupted with AWGN}
	\label{fig:ch03fig07}
\end{figure*}

Two outputs may be produced after segmentation refinement (Fig. \ref{fig:ch03fig02}), a high-resolution point cloud model obtained directly from the interpolated image, and a down-scaled version of the high-resolution segmentation. The surface mesh model obtained from the point cloud will be referred as Point Cloud (PC) model. Since image refinement is performed independently in each image, the PC model is produce directly from a set of equally spaced contours by tiling the cloud points. The surface mesh of the down-scaled version was generated in Simpleware ScanIP$^{TM}$ $v4.0$, with standard pre-smoothing and mesh refinement settings \citep{ScanIPReferenceGuide}\citep{young2008efficient}, and will be referred simply as ScanIP Mesh (SM) model. The segmentation pipeline was tested in noise-free and noisy images, corrupted with AWGN with a standard deviation of $\sigma_{Noise}=10$ HU. Fig. \ref{fig:ch03fig07} shows the sub-stpng of the segmentation refinement process when applied to both noise-free and noisy images. The Hausdorff Distance (HD) and the Mean Symmetric Distance (MSD) \citep{preim2007visualization} were applied to evaluate each segmentation outcome.

\begin{figure}[htb]
	\centering
	\subfigure[]{
	      \label{fig:figure4_7_1}
	      \includegraphics[width = 27.0 mm]{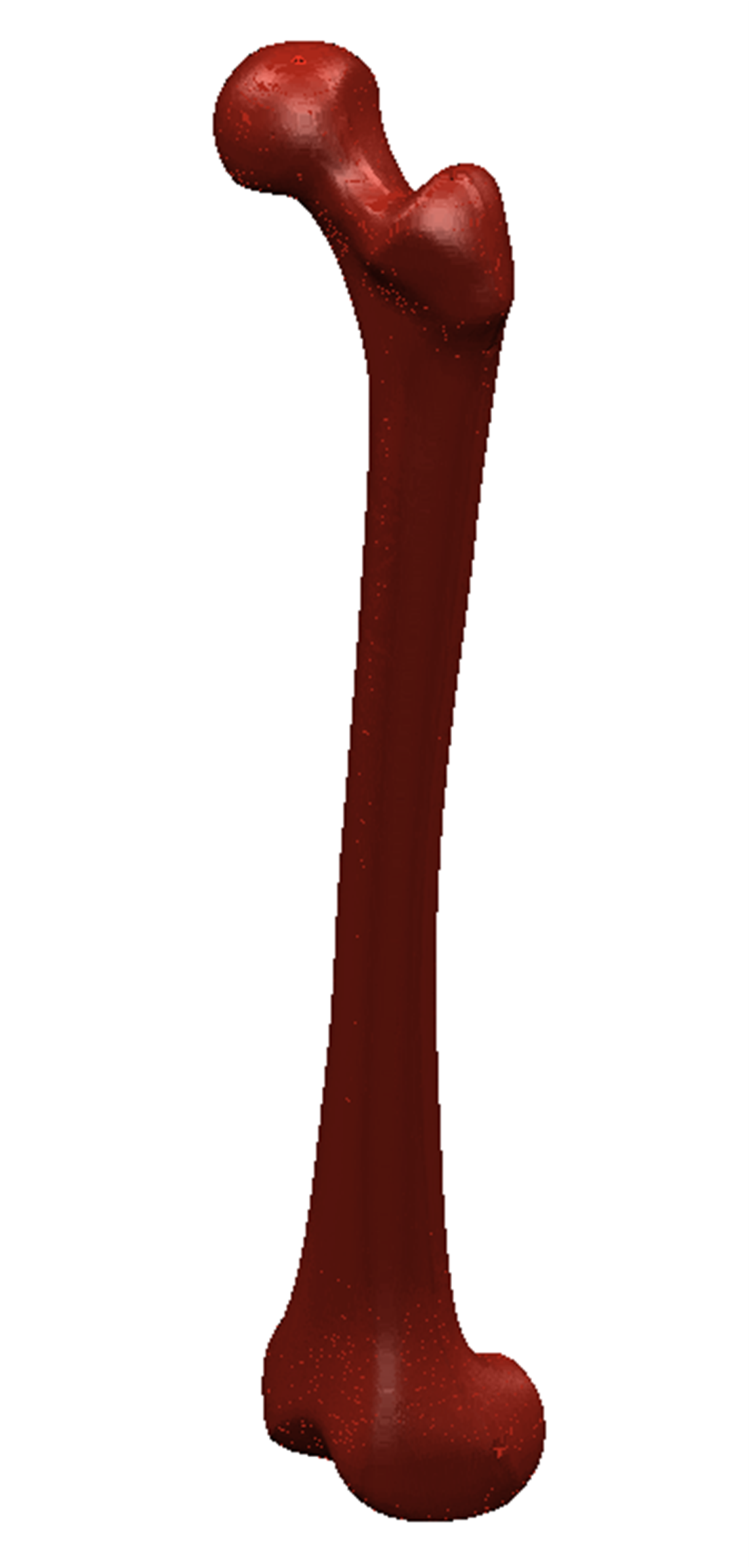}}
	\subfigure[]{
	      \label{fig:figure4_7_2}
	      \includegraphics[width = 27.0 mm]{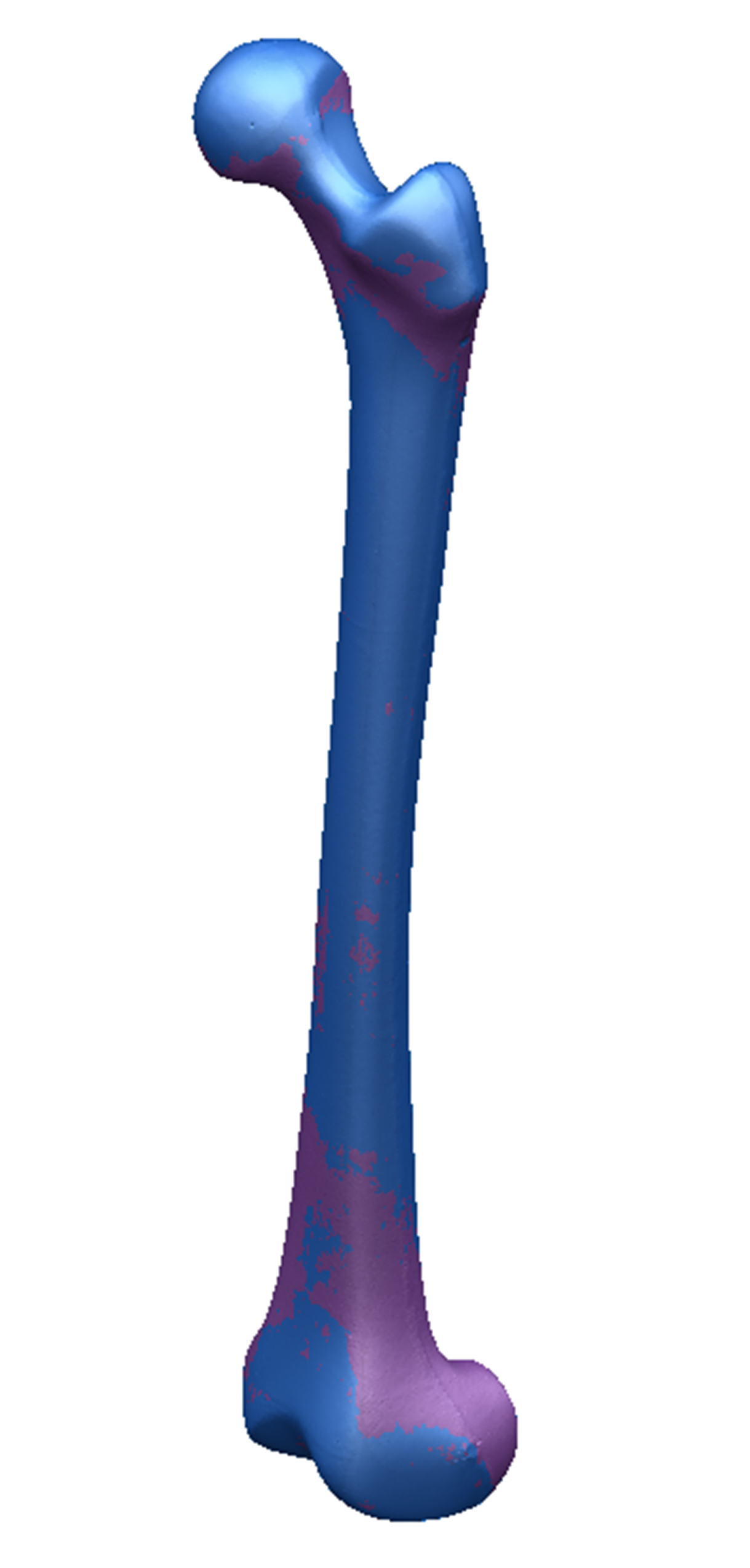}}
	\subfigure[]{
	      \label{fig:figure4_7_3}
	      \includegraphics[width = 27.0 mm]{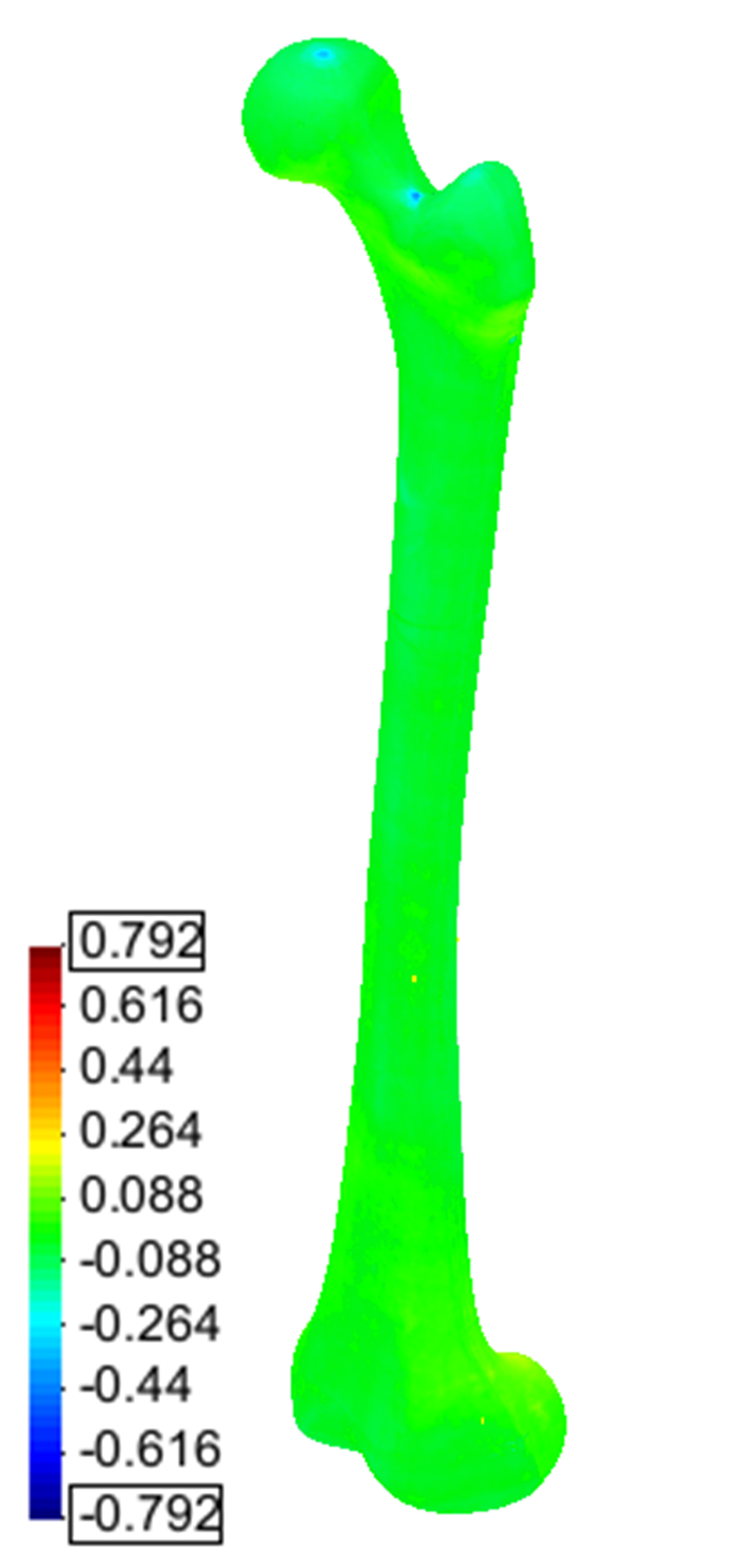}}			
	\caption{In (a) the point cloud obtained from the segmentation of the phantom femur from Dataset \#1, in (b) the surface mesh generated from the point cloud (magenta) and reference femoral surface obtained with the Nikon Metris $LK$ $V20$ (blue), and in (c) the comparison between the two surfaces}
	\label{fig:ch03fig08}
\end{figure}

\begin{figure}[htb]
	\centering
	\subfigure[]{
	      \label{fig:figure4_8_1}
	      \includegraphics[width = 27.0 mm]{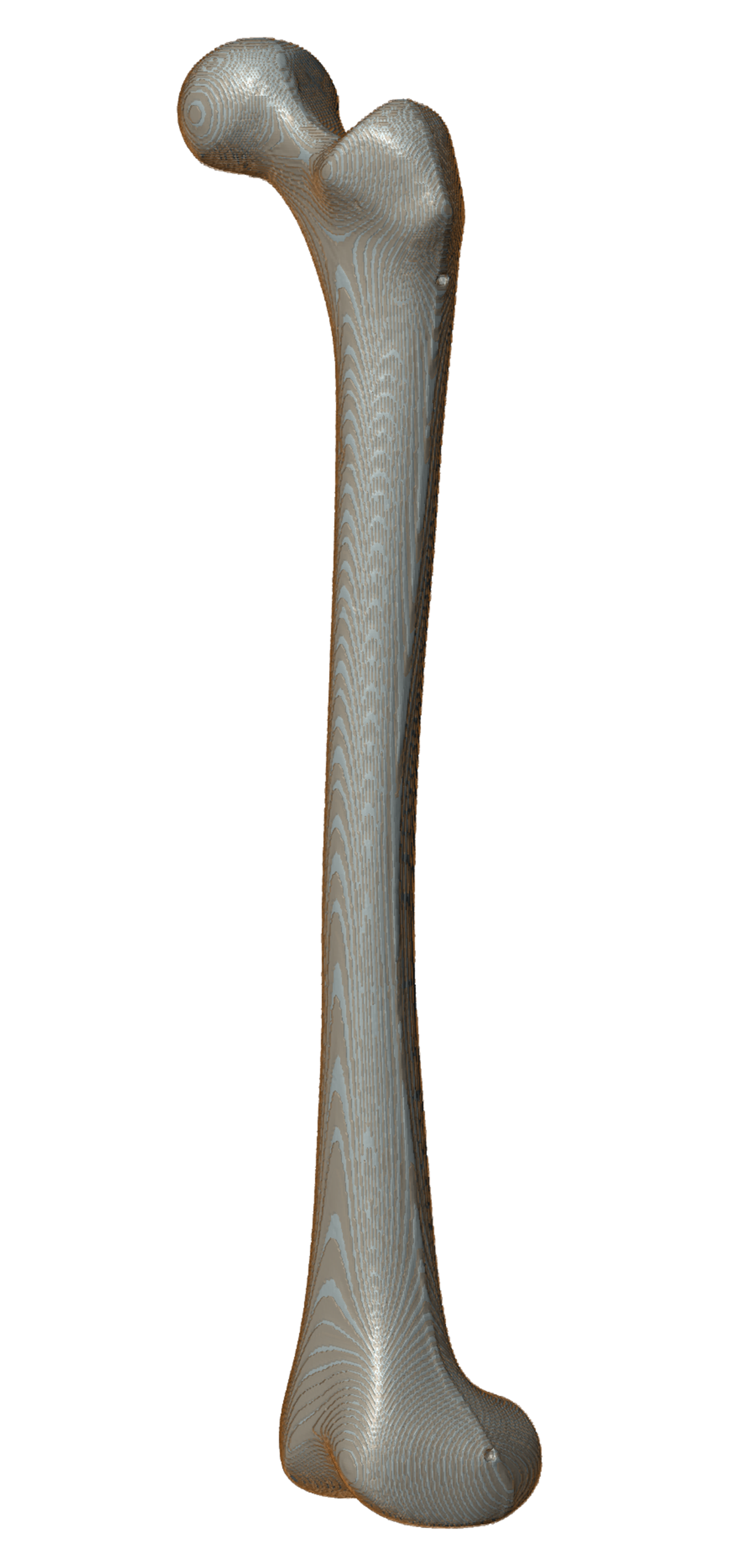}}
	\subfigure[]{
	      \label{fig:figure4_8_2}
	      \includegraphics[width = 27.0 mm]{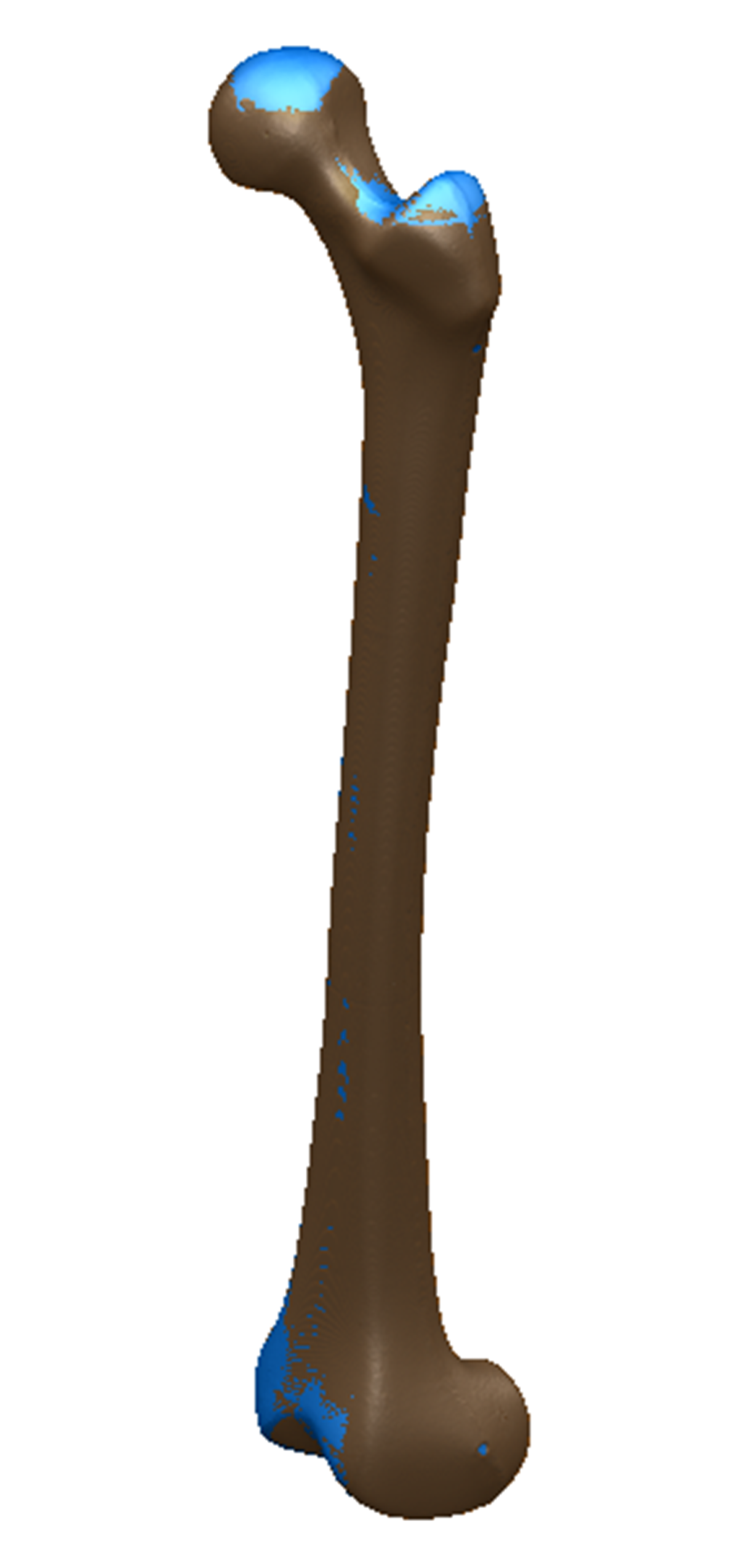}}
	\subfigure[]{
	      \label{fig:figure4_8_3}
	      \includegraphics[width = 27.0 mm]{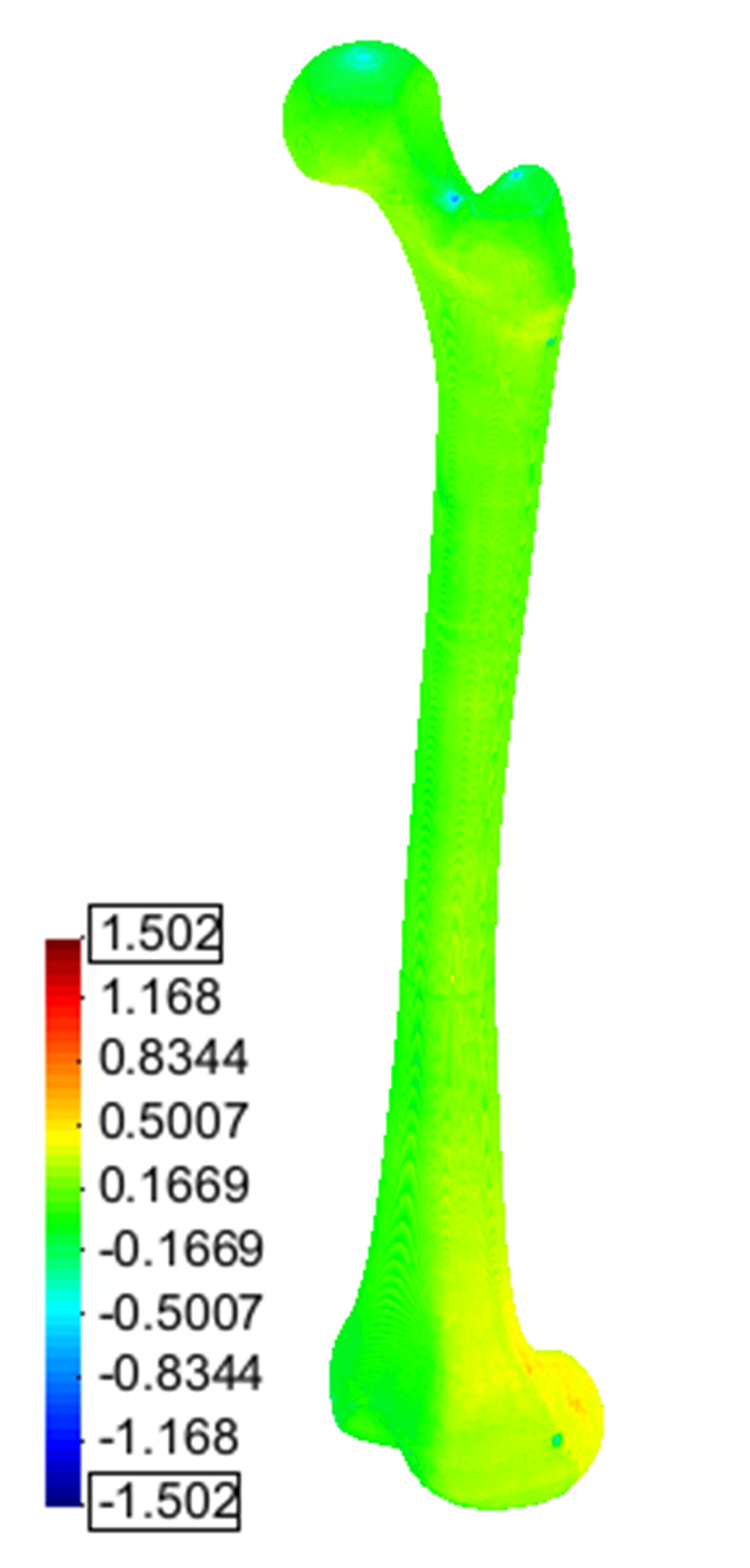}}			
	\caption{On left (a) the domain pre-segmentation and final segmentation after refinement for the Dataset \#1, in the middle (b) the surface mesh generated with Simpleware ScanIP$^{TM}$ and the reference model, and the comparison between the two models (c)}
	\label{fig:ch03fig09}
\end{figure}

Fig. \ref{fig:ch03fig08} shows the final PC model obtained from the segmentation of Dataset \#1 and the comparison with the Nikon Metris $LK$ $V20$ gold standard, whereas Fig. \ref{fig:ch03fig09} depicts the down-scaled segmentation within Simpleware ScanIP$^{TM}$ $v4.0$ overlapped with reference surface. A close agreement between both PC and SM, and the gold standard is observed for both models. The agreement between the PC and SM model and the gold standard for all noise-free and noisy Datasets regarding the distance measures are summarized in Table \ref{tab:ch03Tab02} and Table \ref{tab:ch03Tab03}, respectively.

\begin{table}[htb]
\linespread{1.0} \selectfont \centering
	\caption{Segmentation accuracy for the PC model according to the dissimilarity measures HD and MSD for the noisy and noise-free Datasets}
	\resizebox{88.0 mm}{!}{
	\begin{tabular}{cccccccccccc} \hline \hline
						 & \multicolumn{3}{c}{Resolution, (mm)}          & \multicolumn{3}{c}{Model Deviation, (mm)} \\ \hline 
	Dataset    & \multicolumn{2}{c}{In-plane} & Cross-plane    & Average       & Std.       & Maximum      \\ \hline 
	\#1        & 0.243         & 0.243        & 0.30           & 0.077         & 0.075      & 0.792        \\
	\#2        & 0.525         & 0.525        & 0.30           & 0.119         & 0.109      & 0.841        \\
	\#3        & 0.525         & 0.525        & 3.00           & 0.103         & 0.173      & 2.691        \\
	\#4        & 0.486         & 0.486        & 0.50           & 0.080         & 0.084      & 1.009        \\
	\#5        & 0.972         & 0.972        & 1.00           & 0.159         & 0.164      & 1.073 			 \\ \hline	
	\#1 + AWGN & 0.243         & 0.243        & 0.30           & 0.086         & 0.109      & 0.971        \\ 
	\#2 + AWGN & 0.525         & 0.525        & 0.30           & 0.121         & 0.112      & 0.908        \\
	\#3 + AWGN & 0.525         & 0.525        & 3.00           & 0.178         & 0.170      & 2.187        \\
	\#4 + AWGN & 0.486         & 0.486        & 0.50           & 0.103         & 0.083      & 0.907        \\
	\#5 + AWGN & 0.972         & 0.972        & 1.00           & 0.153         & 0.134      & 1.199  			 \\ \hline \hline
	\end{tabular}}
	\label{tab:ch03Tab02}
\end{table}

\begin{table}[htb]
\linespread{1.0} \selectfont \centering
	\caption{Segmentation accuracy for the SM model according to the dissimilarity measures HD and MSD for the noisy and noise-free Datasets}
	\resizebox{88.0 mm}{!}{
	\begin{tabular}{cccccccccccc} \hline \hline
						 & \multicolumn{3}{c}{Resolution, (mm)}          & \multicolumn{3}{c}{Model Deviation, (mm)} \\ \hline 
	Dataset    & \multicolumn{2}{c}{In-plane} & Cross-plane    & Average       & Std.       & Maximum      \\ \hline 
	\#1        & 0.243         & 0.243        & 0.30           & 0.151         & 0.125      & 0.899        \\
	\#2        & 0.525         & 0.525        & 0.30           & 0.348         & 0.209      & 1.020        \\
	\#3        & 0.525         & 0.525        & 3.00           & 0.262         & 0.192      & 1.810        \\
	\#4        & 0.486         & 0.486        & 0.50           & 0.284         & 0.115      & 1.022        \\
	\#5        & 0.972         & 0.972        & 1.00           & 0.617         & 0.501      & 2.349        \\ \hline
	\#1 + AWGN & 0.243         & 0.243        & 0.30           & 0.172         & 0.130      & 1.164        \\
	\#2 + AWGN & 0.525         & 0.525        & 0.30           & 0.368         & 0.173      & 1.412        \\
	\#3 + AWGN & 0.525         & 0.525        & 3.00           & 0.342         & 0.199      & 1.821        \\
	\#4 + AWGN & 0.486         & 0.486        & 0.50           & 0.273         & 0.174      & 1.550        \\
	\#5 + AWGN & 0.972         & 0.972        & 1.00           & 0.572         & 0.275      & 1.776 			 \\ \hline \hline
	\end{tabular}}
	\label{tab:ch03Tab03}
\end{table}

The average error between the gold-standard surface mesh obtained with Nikon Metris $LK$ $V20$ and the PC model for DS \#1 was $0.077 \pm 0.075$ mm (Fig. \ref{fig:ch03fig08}), whereas for the SM model the average error was $ 0.151 \pm 0.125$ mm (Fig. \ref{fig:ch03fig09}). For the Dataset \#2 the average error of the SM was $0.348 \pm 0.209$ mm, for the Dataset \#3 the average error was $0.262 \pm 0.192$ mm, $0.284 \pm 0.115$ mm for the Dataset \#4, and $0.617 \pm 0.501$ mm for Dataset \#5. Comparing the results of Table \ref{tab:ch03Tab02} and Table \ref{tab:ch03Tab03}, the surface meshes generated directly from the high-resolution point cloud allow the production of more accurate models for all Datasets. The reduction on the average error ranges from 51\% for Dataset \#1 to 26\% for Dataset \#5 ($0.172 \pm 0.159$ mm) for noise-free images. Similar values were observed for the noisy Datasets, and therefore the algorithm shows robustness against image noise.

In the PC surface meshes are triangulated from the 3D points, and model smoothness is directly defined by the smoothness of the level-set contour. Contrariwise, in the SM models the final accuracy is not only dependent on the segmentation accuracy, but also on the settings used for surface mesh generation. The results show that surface mesh tiling allows the accurate definition of the femoral outer surface not only in high-resolution scans, but also in scans with more \textit{ \textquotedblleft clinical \textquotedblright } settings such as with Dataset \#3. Surface meshing directly from the point cloud data effectively avoids the straircase artifacts commonly observed with larger slice thicknesses, and greatly improves the accuracy of the final model.

\begin{figure}[htb]
	\begin{center}
		\includegraphics[width=88.0 mm]{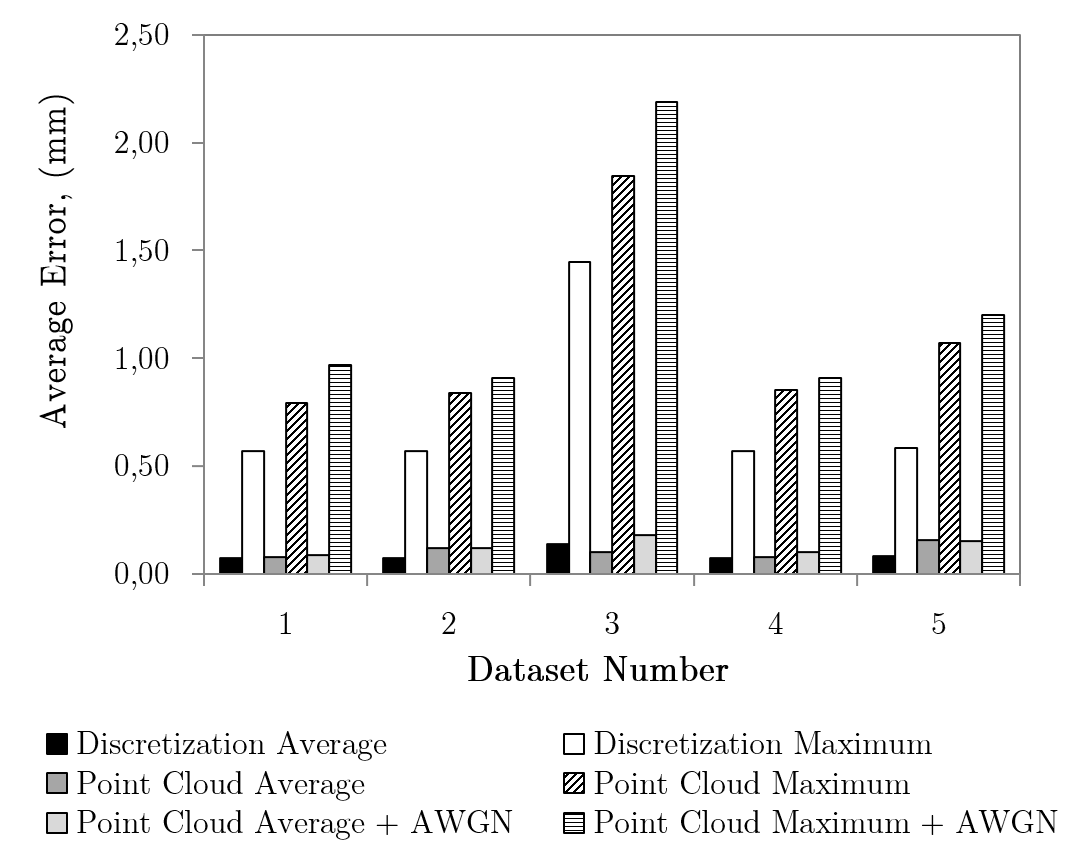}				
		\caption{Comparison between the domain average and maximum error imposed by domain discretization and the average and maximum error of the final model}
		\label{fig:ch03fig11}
	\end{center}
\end{figure}

In section \ref{secIV} for a domain sampling consistent with the FWHM an average error of $0.073 \pm 0.066$ mm was expected. Predictably image segmentation is an additional source of model errors. Nevertheless, the average error obtained for Dataset \#1 is in close agreement with the theoretical discretization error. The results show that the whole chain of image deconvolution, interpolation and segmentation only adds an error of approximately 0.36\% to the theoretical average error. A FWHM of 0.88 mm causes obvious limitations in some Datasets, and the spatial limitations imposed by the PSF superimpose to the pixels size. Fig. \ref{fig:ch03fig11} compares graphically the theoretical average and maximum deviation from the gold standard due to domain discretization considering the FWHM of the PSF, and the average and maximum error of all PC models. Domain discretization contributes in a greater extent to the final deviation from the ground-truth, when compared with the segmentation refinement chain.

\begin{figure}[htb]
	\begin{center}
		\includegraphics[width=88.0 mm]{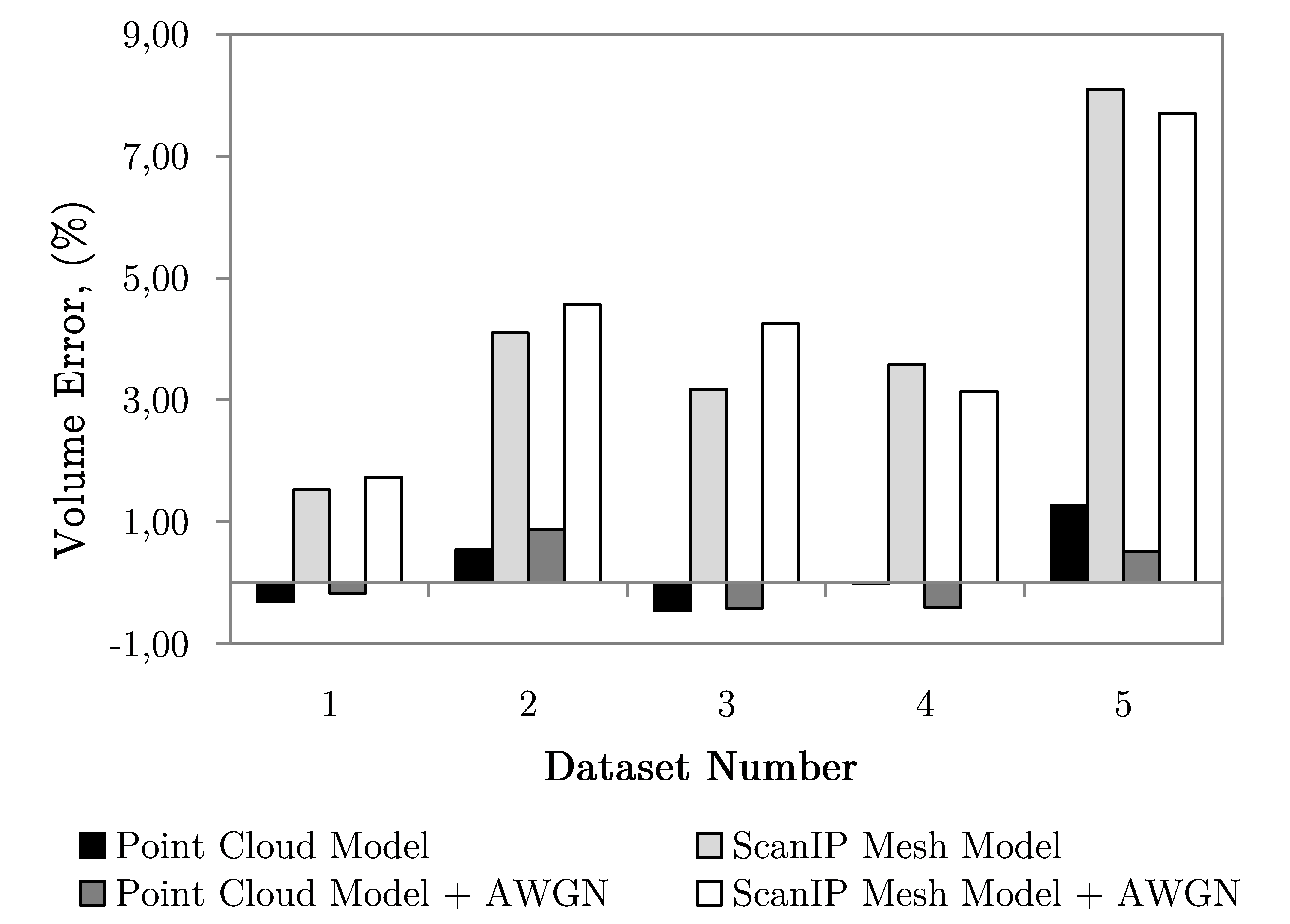}				
		\caption{Volume dissimilarities between the gold standard model obtained with the Nikon Metris $LK$ $V20$, the point cloud-based model and SM model for noisy and noise-free images}
		\label{fig:ch03fig12}
	\end{center}
\end{figure}

Fig. \ref{fig:ch03fig12} shows volume deviation for each final model. The maximum volume deviation was found for the noise free Dataset \#5. The PC models underestimate the ROI in Datasets \#1, \#3 and \#4, more precisely -0.31\%, -0.55\% and -0.01\%, respectively, and overestimate it in 0.55\% in Dataset \#2, and in 1.27\% in Dataset \#5. For the down-scaled model volume overestimation ranges from 1.52\% in Dataset \#1 to 8.09\% in Dataset \#5 for noise free images, and from 1.73\% to 7.70\% for Dataset \#1 to \#5, respectively. Interestingly, for Dataset \#5 the average error and volume deviation decrease with the addition of Gaussian noise. 

\section{Discussion \label{secVI}}

During image acquisition the CT scanner acts as a low-pass filter, eliminating all the high-spatial frequencies. The sharp transitions between different anatomical regions become unclear, and sometimes barely recognizable. The edge blurring effect often leads to the overestimation of the ROI, and the recovery of the original scene is intrinsically limited by our knowledge about the system's PSF. The ability to precisely determine the PSF is of paramount importance if one aims to extract geometrically accurate models from medical images.

In this work we have tested three different phantom objects and their ability to estimate the system's PSF against the CATPHAN 528, which is the phantom routinely used for quality control. The phantom objects aim to estimate the PSF either directly from the system's response to an impulse signal, or indirectly through the estimation of the ESF. Relatively good estimates were obtained with all phantoms. However, the brass wire with 0.10 mm of diameter provided the most accurate estimate of the PSF with a standard deviation error of only 2.9\% when compared with the PSF obtained with the CATPHAN 528. Similar values for the Toshiba $Aquilion^{TM}$ $64$ system were obtained in \citep{verbist2008evaluation}.

Unlike direct measurements, the estimation of the PSF through the analysis of the system's response to strong edges implies additional signal processing stpng. Gaussian filtering has been associated with edge bias by several authors \citep{verbeek1994location}\citep{mendoncca2004bias}\citep{bouma2005correction}. In addition, the ESF may be affected by the presence of image noise during the calculation of the gradient of the attenuation profile. Noise must be eliminated prior to the calculation of the gradient. In this context, \citep{mori2009deriving} proposed an additional blurring correction step to eliminate the Gaussian smoothing effect from the estimate of the MTF estimate. Instead of correcting the result, wavelet thresholding can be used to avoid linear signal filtering. Our results show that although less accurate than the wire phantom, the estimate of the PSF through the ESF and wavelet denoising produces quite good approximations to the real CT spatial blurring function. The wavelet thresholding provides the means to effectively reduce the noise level along the attenuation profile, and to accurately estimate of the ESF without the need for any compensation procedure.

In practice, the accurate quantification of the system's PSF allow us to accurately recover the original image using more standard image deconvolution algorithms. In fact, the accurate estimate of this quantity proved to be quite important to obtain accurate segmentations from image data. In \citep{pakdel2012generalized} it was concluded that reversing the blurring degradation prior to segmentation is essential for the construction of accurate FE models from medical imaging.

Regarding image segmentation, a two-step segmentation pipeline was proposed. The segmentation outcomes were validated through a phantom study, where the final models were compared against a gold standard surface mesh. The gold-standard representation of the phantom object was acquired with a coordinate measuring machine Nikon Metris $LK$ $V20$ with a digital line scanner $LC60-D$, which guarantees 28 $\mu m$ of spatial accuracy.

In the results obtained the maximum average deviation from the gold standard was 0.172 mm for Dataset \#5, whereas the maximum deviation was 2.691 mm for all sets of images considered in the analysis. These results show that the accurate estimation of the system's PSF, conjugated with image interpolation and level-set segmentation provide quite good segmentations of the target geometry. In addition, results provide clear evidences that surface meshes computed from a high-resolution 3D point cloud provide a more accurate representation of the bone than the standard surface meshing from the voxelized data. The point cloud data allows the production of surface models that are both smooth and relatively independent from the surface meshing. Surface triangulation from adjacent contours (or surface tiling) also avoids staircase artifacts due to large section variation between consecutive slices caused by large slice thicknesses \citep{keppel1975approximating}. The advantage of using surface tiling is observable when comparing the results obtained in Datasets \#2 and \#3. The average errors obtained with Dataset \#2 and \#3 differ only in 0.003 mm. Nevertheless, the largest maximum deviation observed in the PCM models may also be a consequence of surface mesh tiling, as can be observed in Fig. \ref{fig:ch03fig08}.

The results show that segmentation accuracy is more dependent on the reconstruction FOV than on the slice thickness. This observation in in agreement with observations found in other phantom studies \citep{ohkubo2008imaging}\citep{meurer2013influence}. Using the high-resolution point cloud to generate the surface meshes may allow the acquisition of image data with more clinical settings, reducing the exposure of the patient to the radiation without a significant loss in model accuracy. In addition and unlike previous works \citep{yao2005estimation}, our results also show that bone segmentation through the level-set method is accurate and possesses numerous advantages against other segmentation methods. The level-set method is robust to noise, it guarantees contour smoothness, is topologically flexible, and allows a straightforward incorporation of user-defined spatial constraints.

Considering the errors associated with domain discretization, the errors produced by the segmentation pipeline varied from 0.41\% to 8.94\% of the final error. These errors are in the same order of magnitude as the discretization itself. In section IV for a domain sampling consistent with the system's PSF the normalized average error is approximately 8.3\% of the sampling or ($0.073 \pm 0.066$ mm).

Domain discretization may also explain the average errors obtained with Dataset \#1 to \#4, which are very similar despite of the voxel size, and the considerable increase in the average error observed for Dataset \#5. In Dataset \#1 and \#4 the system's PSF is the main limitation to the spatial accuracy of the model, whereas in Dataset \#5 the main limitation is the voxel size itself. The average error obtained from Dataset \#1 shows that the whole chain of image deconvolution, image up-scaling through cubic spline interpolation, and subsequent segmentation, there is almost no information degradation or distortion due to the image processing pipeline. Our observations also corroborate previous observations in \citep{cortez20133d} that the domain discretization contributes in a larger extent to the final error of the surface mesh model.

The segmentation in two stpng proposed in this work is very similar to the other methods proposed in the literature. For comparison purposes, the main results obtained by different authors in the literature are summarized in Table \ref{tab:ch03Tab05}. Three studies that have an average error smaller than the maximum average error obtained with the proposed segmentation refinement protocol, namely the optimized 600 HU threshold proposed in \citep{aamodt1999determination}, the Levenberg-Marquardt-based algorithm proposed in \citep{treece2010high}, and the gradient descent algorithm proposed in \citep{pakdel2012generalized}. Nevertheless, in \citep{aamodt1999determination} the average error varied from -0.20 mm to 0.20 mm in the eight femurs considered, which is larger than the maximum average error obtained with our algorithm for all Datasets. The high standard deviation (0.77 mm) shows that the algorithm proposed in \citep{treece2010high} mainly oscillate around the true surface of the cortical bone. In \citep{pakdel2012generalized} slightly less oscilating results were obtained. However, for a comparable spatial resolution (Dataset \#4), the results in this work have a smaller average error of 0.080 mm (with 68.2\% of the surface points within the interval of -0.004 to 0.164 mm) for noise free data, and with an average error of 0.103 mm (with 68.2\% of the surface points within the interval of  0.020 to 0.185 mm) for noisy data.

\begin{table}[htb]
\linespread{1.0}\selectfont
	\centering
	\caption{Bone segmentation accuracy obtained in other studies found in the literature}
	\resizebox{88.0 mm}{!}{
	\begin{tabular}{ccccccc} \hline \hline
													& \multicolumn{3}{c}{Resolution, (mm)}         & \multicolumn{3}{c}{Model Deviation, (mm)} \\ \hline
													& \multicolumn{2}{c}{In-plane} & Cross-plane   & Average     & Std.          & Maximum     \\ \hline 
	\citep{aamodt1999determination}   & 0.312         & 0.312        & 2.00          & 0.03        & 0.52          & 5.75        \\
	\citep{kang2003new}      					& 0.200         & 0.200        & 0.50          & 0.20        & 0.09-0.13     & 0.26        \\
	\citep{oka2009accuracy}       		& 0.293         & 0.293        & 0.63          & 0.46        & 0.03          & 0.49        \\
	\citep{wang2009precision}      		& 0.195         & 0.195        & 1.00          & 0.21        & 0.12          & 0.47        \\
	\citep{liang2010comparative}      & 0.480         & 0.480        & 0.50          & 0.14        & \_\_          & 1.81        \\
	\citep{treece2010high}            & 0.589         & 0.589        & 1.00          & 0.02        & 0.77          & \_\_        \\
	\citep{rathnayaka2011effects}     & 0.390         & 0.390        & 0.50          & 0.18        & 0.20          & \_\_        \\
	\citep{pakdel2012generalized}     & 0.480         & 0.480        & 1.00          & 0.14        & 0.11          & \_\_        \\ \hline \hline
	\end{tabular}}
	\label{tab:ch03Tab05}
\end{table}

The practical implications of an average deviation from the gold standard of 0.172 mm in custom implant modelling are still unclear. In an early study, \citep{carlsson1988implant} found that implant-bone surface gaps of 0.35 mm or more were not bridged by cortical bone, and that this value is close to the critical gap width for which direct lamellar bone apposition occurs onto unloaded implants. In an animal study, \citep{pazzaglia1998relevance} found that there was no bone integration at the interface of roughened titanium rods with 0.30 mm of diametrical gaps at the bone-implant interface. The average error obtained may have minor practical implications in the performance of the custom implants developed from the 3D models for both orthopaedic and trauma applications.In \citep{aamodt1999determination} an inward offset of 0.50 mm to the contours obtained with the 600 HU contour was proposed to avoid custom femoral stem over dimensioning. Endosteal contour shrinkage is needed due to the large standard deviation (0.52 mm) around the average error (Table \ref{tab:ch03Tab05}). Traditional surgical techniques may play a more relevant role in the implant outcome than the geometrical accuracy obtained during implant modelling (see for instance \citep{paul1992development}).

The main limitations of the present study are directly linked with the limitations often associated with every phantom study. Phantom studies are flexible to parameterize and allow the simulation of different acquisition and reconstruction protocols, noise levels, among others. In this type of studies the ground truth may be accurately characterized, and does not change under different environmental conditions. Nevertheless, phantom studies may not be realistic enough to model the complexity of the real data. The CT images used in this analysis have good contrast and are almost noise free. Such images are very difficult or even impossible to obtain in practice. The ACWE are very robust to noise but sensitive to image contrast, therefore in real data the performance of the present algorithm may be slightly degraded. Furthermore, although the use of ACWE enables some degree of flexibility in the pre-segmentation step. There is still however some dependency on the curve initialization. Image under-segmentation seems to be more problematic than over-segmentation, hence in ambiguous pixels it is better to over-segment the domain rather than excluding the pixels. The elimination of such a dependency and the extension of the protocol to the third dimension, in order to obtain a 3D point cloud instead of the current 2D contours are natural evolutions to the current algorithm, and will be considered in future work. These may facilitate the pre-segmentation step and may help to overcome the limitations of surface mesh tiling near bone ends, and allow the production of more accurate models near articulating surfaces.

\section{Conclusions \label{secVII}}

In this work a two-step segmentation pipeline is proposed for accurate bone segmentation from CT image data. The proposed methodology handles segmentation variability by allowing a first free pre-segmentation step, where the user can employ the necessary means to obtain a good approximation to the target segmentation. The second step is standardized and fully automatic and encompasses image restoration, cropping, interpolation, and level-set segmentation. The proposed methodology produces accurate estimates of the target geometries with a maximum average deviation of 0.172 mm. Our results show that surface meshes extracted directly from the high-resolution point cloud reflect more accurately the target ROI. For these meshes the final model accuracy is mostly affected by the image acquisition and reconstruction, rather than by the image segmentation and surface meshing processes. In addition, the CT machine PSF can be accurately determined using a brass alloy wire phantom with 0.10 mm of diameter. The direct measures provide slightly more accurate estimates of the system's Point Spread Function, when compared with indirect measures based on the Edge Spread Function.
\hfill
\hfill

\section*{Acknowledgment}
The first author would like to acknowledge FCT – Funda\c{c}\~{a}o para a Ci\^{e}ncia e Tecnologia (Portugal) for the PhD grant SFRH/BDE/51143/2010. The authors also would like to acknowledge Hospital CUF, Porto (Portugal), Cl\'{i}nica Dr. Campos Costa, Porto (Portugal), and ISQ – Instituto de Soldadura e Qualidade for all technical support provided during this work.

\bibliographystyle{IEEEtran}


\end{document}